\documentclass[pdflatex,sn-mathphys-num]{sn-jnl}

\usepackage{graphicx}%
\usepackage{multirow}%
\usepackage{amsmath,amssymb,amsfonts}%
\usepackage{amsthm}%
\usepackage{mathrsfs}%
\usepackage[title]{appendix}%
\usepackage{xcolor}%
\usepackage{textcomp}%
\usepackage{manyfoot}%
\usepackage{booktabs}%
\usepackage{algorithm}%
\usepackage{algorithmicx}%
\usepackage{algpseudocode}%
\usepackage{listings}%
\usepackage{subfigure}

\begin{document}

\title[Article Title]{Tunable extraordinary optical transmission for integrated photonics}

\author[1]{\fnm{Hira} \sur{Asif}}\email{hiraasif901@gmail.com}

\author*[2]{\fnm{Ramazan} \sur{Sahin}}\email{rsahin@itu.edu.tr}

\affil*[1,2]{\orgdiv{Department of Physics}, \orgname{Akdeniz University}, \city{Antalya}, \postcode{07058}, \country{Turkey}}
\affil*[1,2]{\orgdiv{Turkiye National Observatory}, \orgname{TUG}, \city{Antalya}, \postcode{07058}, \country{Turkey}}

\abstract{The propagation of light through opaque materials, served by periodic arrays of subwavelength holes, has revolutionized imaging and sensor technology with a breakthrough of extraordinary optical transmission (EOT). The enhanced optical transmission assisted by surface plasmon resonances (SPR) has become the most ingenious phenomenon in the field of light-matter interaction. Active tuning of SPR presents a new and simple way to control spectral features of the EOT signal (without the need to change the geometrical structure of the device). This provides a new possibility to integrate an active EOT device with tunable operational frequencies on a single chip of photonic integrated circuits (PIC)- a new scalable instrument in the optoelectronic industry, and quantum technology for improving subwavelength optical imaging and biomedical sensing. In this review, we discuss the fundamentals of EOT, the role of SPR, and how the active quantum plasmonic control of the EOT device makes it a feasible on-chip electro-optic programmable element for integrated photonics.}


\maketitle

\section{Introduction}\label{sec1}
Until the end of the 20$^{th}$ century, it was well-known that light could not pass through a subwavelength aperture in an opaque screen. According to the classical theory of diffraction given by \textit{Bethe} in 1944, the transmission (T), normalized to hole diameter (d), is scaled as $(d/\lambda)^4$, reduces to zero when the wavelength $\lambda$ of incident light is much larger than the hole diameter \cite{Bethe1944}. However, the advent of nanotechnology made it possible to engineer nanostructures and fabricate arrays of subwavelength holes in metals through focused ion beam milling or electron beam lithography techniques. Nanoscale engineering brought new insights into optical phenomena such as diffraction, transmission, reflection, and scattering of light from nanostructures and instigated the fields of subwavelength optics and nanophotonics. The advancements in nano-optics revolutionized optical sensing and nanoscale imaging techniques and uncovered various fascinating phenomena of light at the nanoscale under light-matter interaction. Among them was the extraordinary optical transmission (EOT).
\begin{figure*}[t!]
    \centering
    \begin{subfigure}
        \centering
        \includegraphics[height=1.75in, width=1.9in]{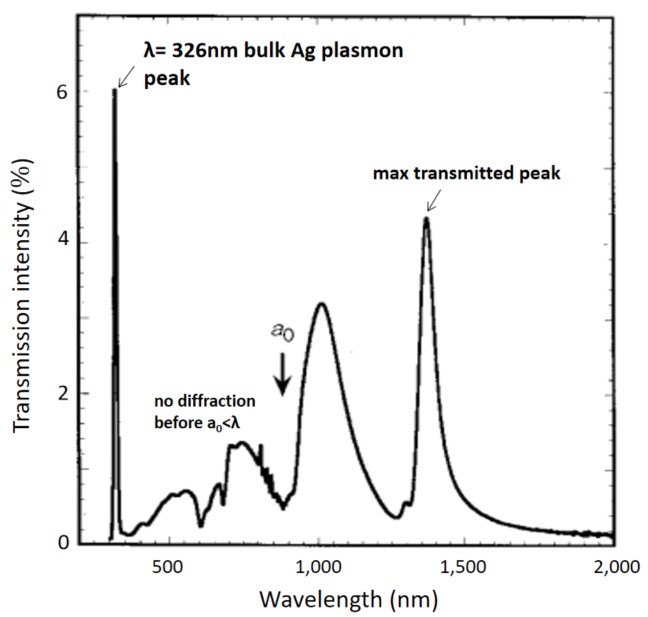}
        \label{fig:1.1}
    \end{subfigure}%
    ~ 
    \begin{subfigure}
        \centering
        \includegraphics[height=1.75in]{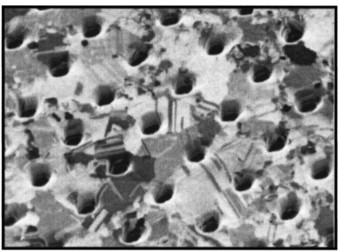}
        \label{fig:1.2}
    \end{subfigure}
    \caption{ (a) Zero-order extraordinary transmission spectrum of an Ag array ($a_o=900$ nm, $t=200$ nm, $d=150$ nm). Reprinted with permission from \cite{Ebbesen1998}. Copyright $1998$ Nature. (b) Focused ion beam image of two-dimensional hole array in a silver film, with hole diameter $d=150$ nm, array periodicity $a_o$=900 nm, and film thickness $t=200$ nm. Reprinted (b) with permission from \cite{Ghaemi1998}. Copyright $1998$ by American Physical Society.}
\end{figure*}\\

When light strikes an optically thick metal film perforated with periodic arrays of subwavelength holes, more light transmits from the metal than impinges on it, known as EOT. In 1998, \textit{Ebbesen} and his co-workers reported this unusual transmission from Ag film, perforated with nanoholes, with transmission efficiency exceeding unity \cite{Ebbesen1998}. To fabricate a hole array structure in Ag metal, the focused ion beam (FIB) was utilized, an image of such a structure is shown in Fig.\ref{fig:1.2}(b), and the zero-order transmission spectrum was obtained with a spectrophotometer using incoherent light sources [see Fig.\ref{fig:1.1}(a)]. Through this experiment, \textit{Ebbesen} observed that EOT is only possible when (i) a TM polarized light is incident perpendicular to the metal film, (ii) the hole diameter is less than the wavelength of the incident light, and (iii) the periodicity of the hole arrays is resonant with transmission wavelength, (iv) solely metal film is utilized instead of semiconductor material such as Ge film \cite{Reuven2008, Ghaemi1998}. Under these conditions, it was found that the transmission was not only higher than the light impinging on the structure, but it also breached the transmission limit defined by Bethe. In addition, transmission efficiency is greater than the area-to-aperture ratio of the film, indicating that light incident on the metal surface in between the holes was also transmitted. Moreover, transmitted light propagates in a collinear direction which makes it intense and beamed rather diffracted \cite{Garcia2010}. With these characteristics features such as high transmissivity, collinearity, and spectral coherence, EOT has gained substantial interest in the research community and has shown great potential in sensing, energy transfer, lasing, optical filtering, and photonic integrated circuits (PIC) \cite{Maier2007, Kabashin2009, Homola1999, Anker2008, Liu2015, Angela2007}.  
In this review, we aim to discuss new developments that have occurred in the tunable EOT phenomenon and mainly focus on how the transition from passive control of the EOT signal occurs towards active tuning of EOT light for its potential implications in PIC. In section \ref{sec2}, the fundamentals of EOT are discussed with a detailed description of the basic mechanism behind the EOT phenomenon. In section \ref{sec3}, we provide a brief overview of passive control of EOT signal through geometrical parameters and the advancement in this area over the past two decades. Section \ref{sec4} focuses on the active control of the EOT device through electrically tunable materials. In section \ref{sec7}, we explore the quantum control of the EOT signal following recent developments to optimize the EOT device as an \textit{electrically-programmable EOT switch}. In section \ref{sec8}, we discuss the potential implications of plasmon-based EOT in PIC chips.  

\section{Fundamentals of Extraordinary Optical Transmission}\label{sec2}
\begin{figure*}[t!]
        \centering
        \includegraphics[height=2.0in]{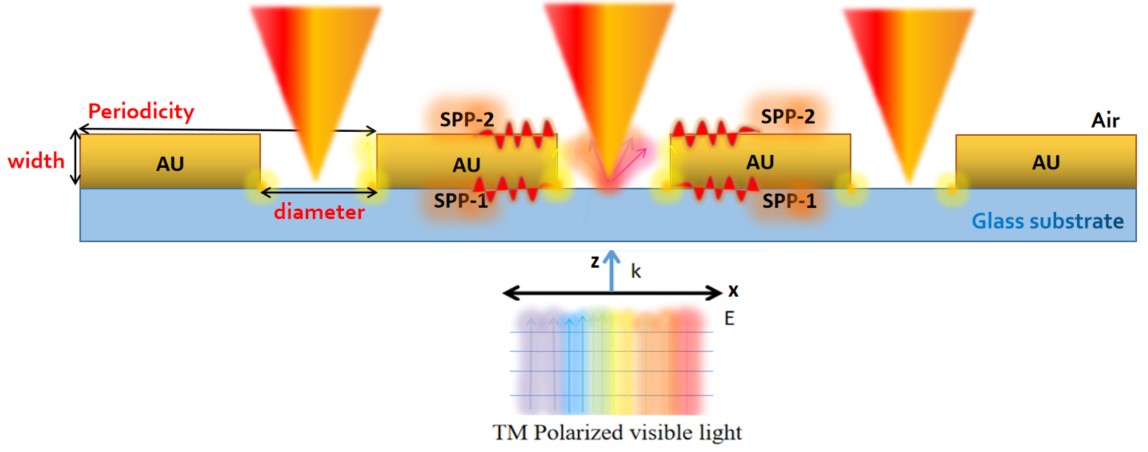}
        \label{fig:2}
    \caption{Schematic diagram of an EOT structure (cross-section view). Au metal-nanohole array layered on a glass substrate. A p-polarized light incident from the glass side excites SPPs. The transmitted light that emerges from the metal-air side is mediated by the coupling of resonant and nonresonant SPP modes at both metal interfaces.}
\end{figure*}

The discovery of EOT has led many scientists to anticipate the mechanism behind the enhanced transmission and this inquisitiveness resulted in many follow-up studies that suggested the main mechanism behind EOT is the excitation of surface plasmon polaritons (SPPs) \cite{Ghaemi1998, Barnes2004}. SPPs excite on the surface of a metal film perforated with subwavelength holes, due to the coupling of incident light to the free electrons of the metal. These SPPs in the form of evanescent waves propagate along the metal-dielectric interface and decay exponentially perpendicular to the surface. In a semi-infinite metal-dielectric media, the dispersion relation of SPPs is characterized as the complex, frequency-dependent wave-vector $k_{spp}$ which is obtained by solving Maxwell's equation with appropriate boundary conditions \cite{Barnes2003}. 
\begin{equation}
    k_{spp}=k_o\sqrt{\frac{\epsilon_m\epsilon_d}{\epsilon_m+\epsilon_d}} 
    \label{eq:1.1}
\end{equation}
Here, $\epsilon_{m} (\omega)$  and $\epsilon_{d}(\omega)$ are the frequency-dependent dielectric permittivity of metal and dielectric medium, respectively. To excite SPPs on the surface, one of the permittivities must have the opposite sign. This condition is satisfied by metals because $\epsilon_{m}$ is both negative and complex. On the other hand, when charge densities in the metal interact with the electromagnetic (EM) field, the momentum of surface plasmon $\hbar k_{spp}$, is greater than the free-space momentum ($\hbar k_o$) of the incoming photon ($k_o$ is a free space wavevector). However, the periodic array of subwavelength holes provides extra momentum to incoming photons through reciprocal lattice vectors, G$_{x}=$G$_{y}=2\pi/p$, as a function of periodicity, $p$, and allows incident light to couple to SPPs. The momentum matching condition for the wavevectors then becomes, $k_{spp}=k_o\pm i$G$_x \pm j$G$_y$, where \textit{i} and \textit{j} are the indices for the scattering orders which determine the direction of SPPs. Using this momentum matching condition in Eq.\ref{eq:1.1}, the wavelength of SPPs, excited by a Bragg's-type scattering over a two-dimensional nanohole array is given by \cite{Sambles1991}, 
\begin{equation}
    \lambda_{spp}=\frac{p}{\sqrt{i+j}}\sqrt{\frac{\epsilon_m\epsilon_d}{\epsilon_m+\epsilon_d}} 
    \label{eq:1.2}
\end{equation}
where $p$ is the period of hole array and $\epsilon_m$ is the real component of complex dielectric function of metal obtain through Lorentz-Drude model. The real part belongs to resonance wavelength whereas the negative imaginary part provides the non-radiative damping effect. From Eq.\ref{eq:1.2} it is clear that the wavelength of SPPs is influenced by the geometrical parameters of the nanostructure such as the periodicity of hole array, properties of the metal film, dielectric background, and the polarization of the incident light. The relation of the EOT phenomenon with SPPs drives from the fact that the wavelength of the transmitted light shows similar dependence on the lattice period, dielectric medium, and the polarization of the incident light. The excitation of SPPs at the metal-glass interface and their constructive interference at both interfaces allow more light to pass through the holes than would be expected based on conventional optics.\\ 
To explicitly understand the scattering of EOT light to the far field through the excitation of SPPs on both interfaces, we delve deeply into the microscopic description of EOT. Fig.\ref{fig:2} demonstrates a cross-section of the Au metal-hole array structure layered on a glass substrate with a thin adhesion layer of Cr. A TM-polarized plane wave, propagating in the z-direction incident on the glass side, excites SPPs at the metal-glass interface which tunnels through the holes.
After passing through the hole, the SPPs interfere with the evanescent waves that are scattered at the metal-air interface resulting in the enhancement of light that propagates to the far field. The role of SPPs in this way is to enhance the fields associated with the evanescent waves, which results in zero-order extraordinary transmission \cite{Barnes2006}. By zero-order, it means that the incident and transmitted light are collinear, supported by the tunneling transition of SPPs modes. This collinearity occurs when the wavelength of incident light is greater than the grating period, which makes the direction of propagation, polarization, and divergence of the transmitted beam identical to the incident beam \cite{Lienau2004}. Apart from bound SPPs modes, surface plasmons are also excited at the rim of the subwavelength hole in the form of localized surface plasmons (LSP). In contrast to SPPs, LSP can be excited through direct illumination of electromagnetic light on the curvature of the hole structure \cite{Degiron2005}. The intense localized modes at the edges of the hole cavity provide directionality to SPPs waves while tunneling through the hole and their coupling with SPPs mediated by Fano resonance contributes to enhanced transmission \cite{Chen2015, Bao2008}. These plasmonic modes (SPPs and LSP) form the basis of enhanced transmission, and their dynamical response at the top and bottom interfaces plays a crucial role in the optical properties of EOT light. Consequently, modulating the spectral, spatial, and temporal dynamics of these modes enables coherent control of the EOT signal. Fundamentally, the spectral features of the SPPs and LSP modes depend on the geometry and dimensions of the metal–hole nanostructure \cite{Sangiao2016}. More specifically, geometrical parameters, such as hole diameter, periodicity, and the dielectric medium alter the response of both plasmon resonances (SPPs/LSP). For a very thin metal film, the tunneling of SPPs through the holes becomes resonant due to their coherent overlapping or interference on the top and bottom surfaces of the film. These diffraction/interference effects are accurately determined by the transmission spectrum. Nevertheless, the peak spectral position of transmitted light is analyzed from Eq.\ref{eq:1.2}. The peaks in the transmission spectrum indicate the resonant wavelength of those SPPs modes which interfere constructively with the outgoing modes and result as peaks in the transmission spectrum.\\
The spectrum of EOT contains very well-defined characteristic features. The bulk plasmon resonance of Ag metal is depicted by the first peak of the EOT spectrum as illustrated in Fig.\ref{fig:1.1}(a). After the wavelength coincides with the periodicity of hole arrays, the spectrum shows two broad peaks of resonant transmission. These peaks are known as transmission maxima, resulting from the SPPs coupling on the same surface as well as both sides of the metal interface \cite{Popov2000}. The peak’s width is proportional to the inhomogeneous broadening of the holes \cite{Van-der-Molen2000} and also reflects the oscillation time of the SPPs modes. The dips along maxima result from Wood’s anomaly that appears for wavelengths diffracted from the perforated structure and propagates tangent to the metal surface. Whereas, the width of the dip is proportional to the incident beam’s dispersion, which is entirely a geometrical effect \cite{Wood1902, Maradudin2016}. 

\section{Passive Control of EOT}\label{sec3}
Following the analysis of the EOT spectrum, the key factors that have a significant impact on the shape and spectral characteristics of the transmission spectrum are metal film thickness, hole diameter, array periodicity, and dielectric medium at the metal interface. Modulating these parameters leads to the passive controlling of the EOT system and these geometrical parameters greatly influence the characteristics of the EOT signal. In the following sections, we discuss the impact of these structural parameters on the enhanced transmission.

\begin{figure}[h]
\centering
\includegraphics[width=0.9\textwidth]{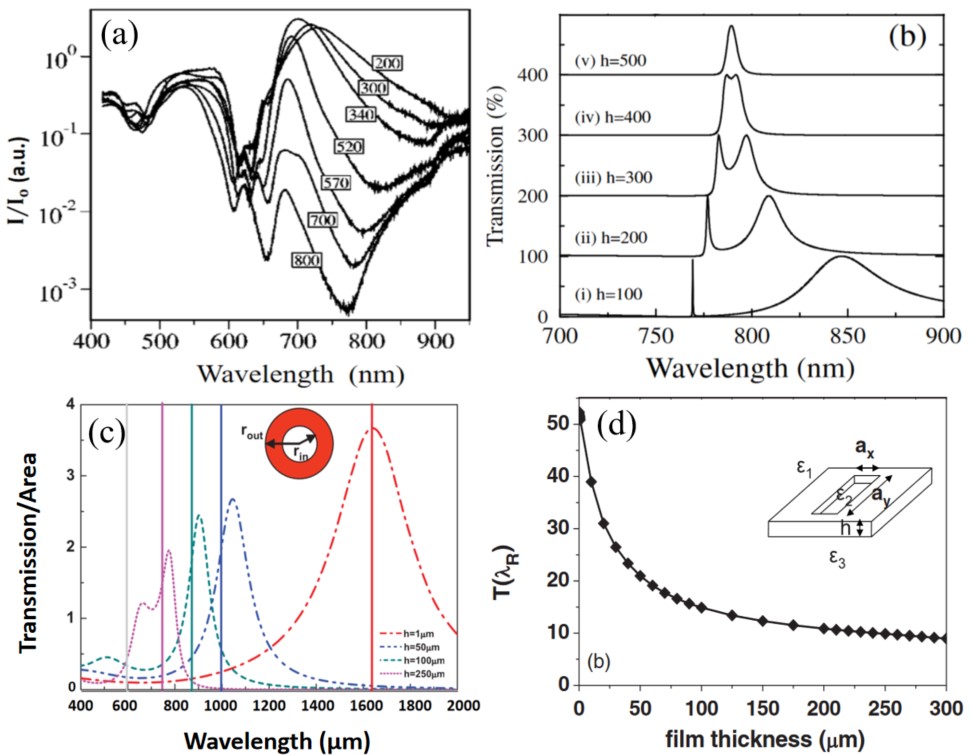}
\caption{(a) Extraordinary transmission spectra from square arrays of subwavelength holes (d=300 nm, p=600 nm) for different values of hole depths h. Reprinted from \cite{Degiron2002}, with the permission of AIP publishing. (b) Transmission for different film thicknesses h in nm, curves shifted by $100\%$. Reprinted (b) with permission from \cite{Martin-Moreno2001}. Copyright 2001 by American Physical Society. (c) Normalized transmission spectra of an isolated annular hole ($r_{in}$ = 90 $\mu$m, $r_{out}$ = 100 $\mu$m) in a PEC for different film thicknesses h with $\varepsilon = 9$ and $\varepsilon_2 = 1$. Vertical lines depict resonant wavelengths $\lambda^{(a)}_{R}$. (d) Normalized transmission intensities at $\lambda_{R}$ for an isolated rectangular hole with $\varepsilon = 12$ and $\varepsilon_2 = 1$. Reprinted (c,d) with permission from \cite{Carretero-Palacios2012}. Copyright (2012) by the American Physical Society.}
\label{fig:3}
\end{figure} 

\subsection{Effect of film thickness/hole depth and the dielectric medium on EOT}\label{subsec1}
The intensity of the transmission peak is decreased by increasing the hole depth (film thickness), investigated by \textit{Degiron et al} in the experimental study for the case of a square array of cylindrical holes. For shallow holes, SPPs on two interfaces coupled via evanescent waves leads to the resonant coupling of SPPs modes which enhances the efficiency of transmission. As the holes get deeper, most of the SPPs modes at both interfaces do not couple efficiently resulting in a weak transmission that falls off exponentially with the increase in the hole depth. As the film thickness decreases the transmission intensity increases exponentially along with a broadening in the peak \cite{Degiron2002} [see Fig.\ref{fig:3}(a)]. \textit{Martin-Moreno et al} explain the broadening of the transmission peak through the minimal model, according to which an increase in the hole depth causes the fast radiative decay of SPPs modes which in turn shortens the lifetime of SPPs oscillations and increases the linewidth of the transmission peak \cite{Martin-Moreno2001} [see Fig.\ref{fig:3}(b)].
On the other hand, \textit{Carretero et al} presented a quantitative theory to investigate the spectral shift in the transmission resonance as a function of film thickness and dielectric environment. In the detailed analysis, the results demonstrated that when the metal film thickness is much smaller than the wavelength of light, the transmission is controlled by effective admittance of vacuum in the hole. When the film thickness is larger than the wavelength, the transmission is controlled by the cutoff of the fundamental waveguide mode inside the hole. In the case of thin film combined with a high-index dielectric, the spectral position red-shifts compared to the cutoff wavelength of the hole, and the transmission efficiency is increased substantially, as shown in Fig.\ref{fig:3}(c) \cite{Carretero-Palacios2012}.
Moreover, the transmission intensity decrease exponentially with the increase in the film thickness, as shown in Fig.\ref{fig:3}(d). \textit{Agrawal et al} also explored the impact of film thickness on the enhanced transmission. They observed transmission for film thickness ranges from $\delta/15$ to $2\delta$, where $\delta$ is the skin depth, which is $\approx$ 25 nm for silver at the optical frequencies. In a detailed quantitative analysis, it was shown that as the film thickness increases from $\delta/15$ to $\delta/6$, there is a discernable enhancement in transmission resonance frequency. For the film thickness equal to the skin depth, there is a sublinear enhancement in transmission. However, no significant enhancement is observed for the films with thicknesses more than twice of skin depth \cite{Agrawal2005}. These studies suggested that optimal film thickness enhances plasmonic coupling and boost transmission efficiency.
The dielectric medium at the interface of metal surfaces also has a significant impact on the transmission efficiency. The change in refractive index n, caused by different dielectric mediums at the bottom and top interfaces of metallic film, effects the coherent coupling of SPPs modes on both surfaces \cite{Krishnan2001}. The coupling becomes stronger for identical dielectric constant of sub- and superstrate.
\begin{figure}[h]
\centering
\includegraphics[width=0.9\textwidth]{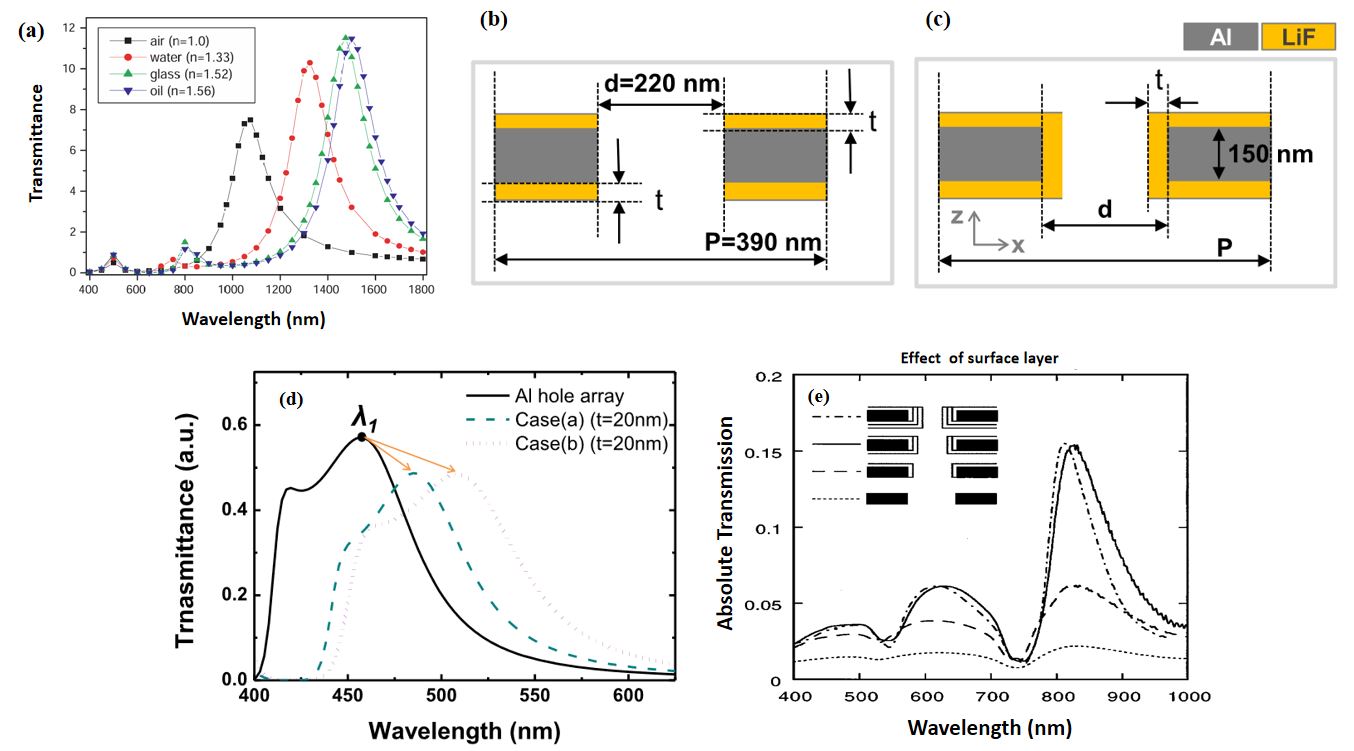}
\caption{(a) Transmittance spectra for thick Au film with a 15 nm wide slit for different materials within the slit. Reprinted with permission from \cite{Lindberg2004}. \copyright Optica Publishing Group. (b,c) Schematic diagram of the free-standing aluminum (Al) films coated with a dielectric material (lithium ﬂuoride, LiF) on the top, bottom and inside the hole of metal. (d) Transmission spectra of Al hole arrays for the cases (a) and (b). Reprinted figures (b-d)with permission of IOP publishing from \cite{Do2014} permission conveyed through Copyright Clearence Center, Inc. (e) Transmission spectra of Ni membrane (t= 250 nm, p=750 nm) coated with Ag film of the thickness (t= 20 nm). Reprinted from \cite{Grupp2000}, with the permission of AIP Publishing.}
\label{fig:4}
\end{figure}
In this way, SPPs resonances at the two interfaces coincide which enhances the transmission by a factor of 10 \cite{Lienau2004}. \textit{Lindberg et al} investigated the dependence of transmission wavelength on the material above and inside the single slit metal film. As the refractive index of slit medium increases, there is significant redshift in the transmission wavelength, as shown in Fig.\ref{fig:4}(a) \cite{Lindberg2004}.
\textit{Yun Seon et al} demonstrate the enhancement in the transmission efficiency along with the shift in its spectral position by adding a thin layer of lithium fluoride (LF) with a refractive index ($n=1.37$) above [Fig.\ref{fig:4}(b)] and inside [Fig.\ref{fig:4}(c)] the metal surface \cite{Do2014}. Due to the high refractive index of LiF as compared to air, more surface charges are bound along the edges of the holes in Fig.\ref{fig:4}(c), which causes a strong dipole moment around the hole.  The localized charges (LSP) around the rims of the holes at both surfaces acts as a Fabry-Perot resonator interacting with evanescent waves. The increase in the interaction between the surface charges results in the lower energy level of surface plasmon due to which transmission peak redshifts from $\lambda_1$ [see Fig.\ref{fig:4}(d)]. \textit{Grupp et al} also suggested that the transmission efficiency depends on the dielectric properties of the in-plane metal surface within the skin depth \cite{Grupp2000}. Fig.\ref{fig:4}(e) shows an increase in the transmission efficiency by adding a 20 nm layer of Ag metal to the Ni membrane perforated with nanohole. The enhancement of EOT in a broken-symmetry structure is obtained by the phase-matched SPPs on both surfaces and the consistency of dielectric around the surface. The consistency of the dielectric medium around all the surfaces is beneficial for the Febry-perot resonant interaction between two LSP modes, which also results in higher transmission \cite{Park2008}. These studies suggested that the resonance wavelength of surface plasmons is strongly influenced by medium around the metal surface and its propagation constant changes drastically as the function of hole thickness. For thin films, resonant coupling of SPPs enhances transmission efficiency, while increased film thickness leads to fast radiative decay of SPPs modes, broadening the transmission peak. Optimal thickness lies within a few multiples of the skin depth of the metal. However, these are not only the counting factors to control the plasmon frequency and eventually EOT dynamics. In the next section, we discuss the impact of hole array periodicity and its shape on SPP propagation and how it influence the transmission resonance from metal-hole array structure.
\subsection{The role of array periodicity and lattice shape on EOT}\label{subsec2}
Subwavelength hole arrays play a major role in the enhancement of extraordinary optical transmission. \textit{Przybilla et al} investigated the extraordinary transmission by comparing a single-hole metal film to an array of subwavelength holes \cite{Przybilla2008}. An isolated hole acts like a scatterer of incident light. This scattering yields excitation of LSP on the hole edges, reflected waves around the hole, evanescent waves inside the cavity, and propagating waves on the surface of the film. In the case of a single hole these SPPs dissipate quickly and do not contribute much to the transmission. However, for arrays of holes, SPPs diffraction from the patterned surface resonantly coupled with other excited modes and enhance the transmission. Furthermore, if the patterned structure is at both interfaces, the diffraction effects on the exit side can lead to a well-collimated beam, and the transmission efficiency is enhanced one to two orders of magnitude in contrast to single hole \cite{Schatz2005}. In the SPPs-based theory, the coupling of incident light to SPPs is supported by the periodicity of the hole lattice. However, the propagation length of SPPs is much smaller than the lattice size which results from the ultrafast damping of SPPs waves. In this way, the translation symmetry of the array plays an important role in the enhanced transmission \cite{Matsui2007, Bravo-Abad2007}. {\textit{Mei et al} investigated the impact of lattice periodicity and its shape on the transmission efficiency and found out that the transmission intensity strongly depends on the design of the lattice period. In contrast to random hole arrays, quasiperiodic hole arrays provide long-range orders and offer some dominant discrete reciprocal vectors which enable SPP-light coupling efficiently and yields enhanced EOT from nanohole arrays.
\begin{figure}[h]
\includegraphics[width=0.9\linewidth]{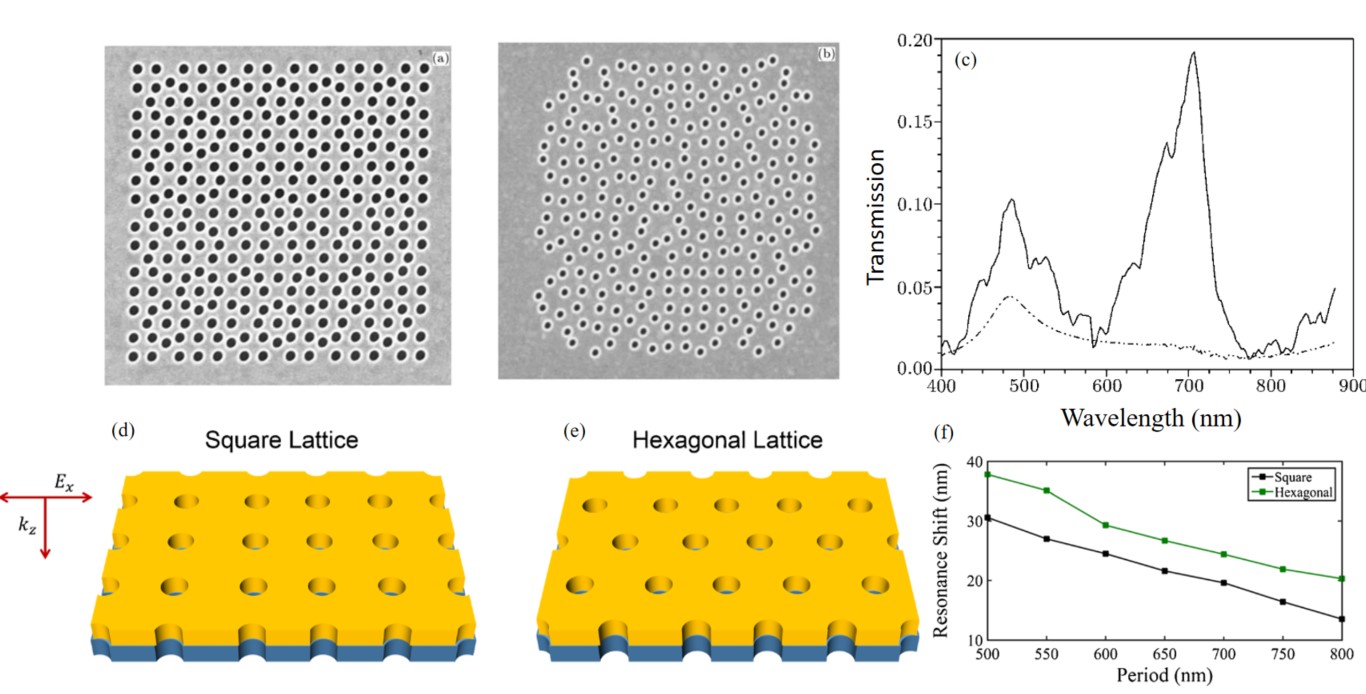} \newline
\centering
\caption{SEM images of (a) quasiperiodic and (b) non-periodic hole arrays in an Au film (t=120 nm). (c) Transmission spectra of quasiperiodic (solid line) and nonperiodic (dashed-dotted line) hole arrays. Reprinted from \cite{Mei2006}\ with the permission from IOP science. (d,e) Square and hexagonal lattice of hole array system in an Au film layered on silicon nitride, respectively. (f) Variation in the spectral position of peak transmission wavelength as a function of lattice period for both lattice structures. Reprinted from \cite{Eksiolu2016}, with permission from Springer Nature.}
\label{fig:5}
\end{figure}
Figure.\ref{fig:5}(a) and \ref{fig:5}(b) show nanohole structure with quasiperiodic and non-periodic lattices perforated in 120 nm thick Au films, respectively. The result demonstrated long-range orders effect in quasiperiodic hole arrays which exhibit about $20\%$ enhanced transmission at 707 nm wavelength. In contrast to this, non-periodic hole arrays with short range orders do not offer such reciprocal lattice which leads to weak coupling of plasmons and no enhancement in transmission peak, as shown in Fig.\ref{fig:5}(c) \cite{Mei2006}. 
In another experimental study, enhanced transmission is found in a random periodic structure with a lack of translation symmetry but long-range orders. To see the impact of lattice shape on the linewidth and spectral dynamics of transmission, \textit{Ekşioğlu et al} investigated the optical response of plasmons from square and hexagonal periodic arrays of circular holes. The schematic of Au film with these lattices layered on a silicon nitride is shown in Fig.\ref{fig:5}(d) and \ref{fig:5}(e). Figure.\ref{fig:5}(f) demonstrate the shift in the resonance wavelength of transmission as a function of lattice period and for both square and hexagonal lattices the transmission peak redshifts as the lattice period increases \cite{Eksiolu2016}.} As the matter of fact, hexagonal arrays supports large figure-of-merit values due to its enhance EOT response. Moreover, nanoholes with the hexagonal lattice show sharper linewidth in contrast to the square lattice. This is due to the fact that a closed paked hexagonal lattice has highest nanohole density and well-defined periodicity for long range SPP resonances \cite{Ohno2016}.
The scattering of light from each hole in the array gives rise to several evanescent waves propagating away from the hole. When these composite evanescent waves arrive at the next hole, it interfere with the light directly incident on the hole. Similarly, composite waves scatter and interfere with SPPs at the other interface \cite{Lezec2004}. The enhancement and suppression in the transmission depends on the relative phase of the interfering waves. Some theoretical studies have confirmed the role of interference of evanescent waves \cite{Sarrazin2003, Scheuten2005} and the role of SPPs surface waves in the transmission process.
In such a structure the transmission maxima occur due to the constructive interference of light that scattered from each hole and diffracted SPP surface waves, which is referred to as interference modal \cite{Pacifici2008, Li2006}. 
\begin{figure}[h]
\includegraphics[width=0.9\linewidth]{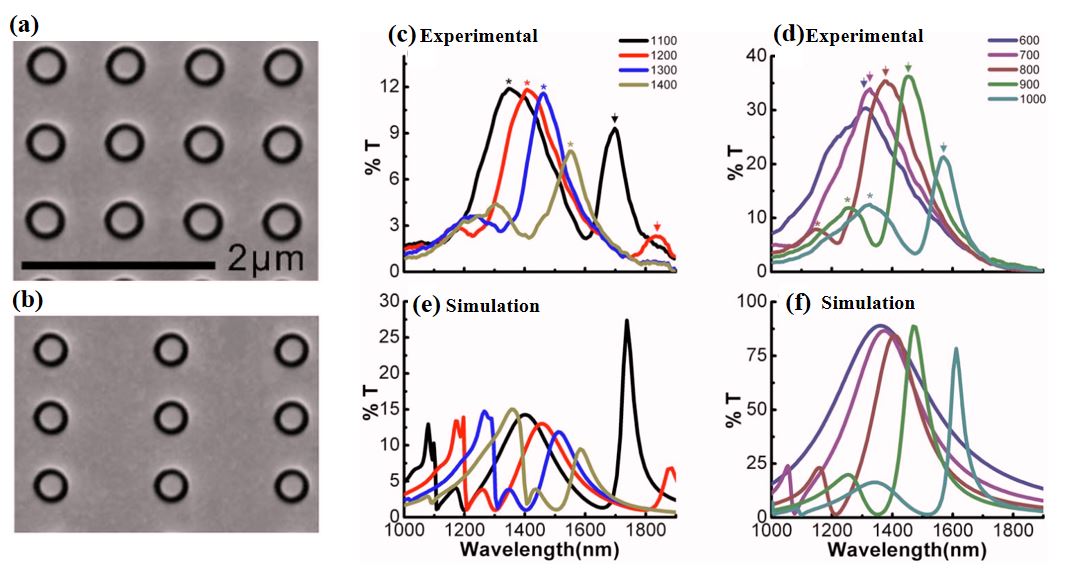} \newline
\centering
\caption{SEM images of FIB milled annular aperture arrays (AAAs) with radius R1 = 125 nm, R2=215 nm, (a) period = 800 nm, (b) AAAs with period = 1400 nm on the x-axis and 800 nm on the y-axis. (c,d) Experimental and (e,f) simulated transmission spectra of AAAs for different periods of 1400-1100 nm and 1000-600 nm corresponding to $\pm$(1,1) Au/quartz and $\pm$(1,0) Au/quartz SPP-Bloch modes, respectively. Reprinted from \cite{Kofke2009}, with the permission from AIP publishing.}
\label{fig:6}
\end{figure}\\
Apart from lattice shape and symmetry, its periodicity also strongly impacts on the peak spectral position of transmission resonance. Following the dispersion relation of surface plasmons the periodicity plays an important role in the excitation of coherent SPP modes on the patterned structure which propagates along the surface. This suggests that the spectral position of the surface plasmon can be modulated as a function of the periodicity through following expression in Eq.\ref{eq:1.2}. To see the impact of array periodicity on the EOT spectrum, \textit{Kofke et al} investigated the annual aperture arrays (AAAs) in a 150 nm thick Au film with two different lattice periods. SEM images of AAAs with periods, 800 nm and 1400 nm along x-axis are shown in Fig.\ref{fig:6}(a) and \ref{fig:6}(b), respectively. Figures.\ref{fig:6}(c) and \ref{fig:6}(d) shows the experimental results with spectral shifts in transmission spectra for array periodicity ranges from 600 nm - 1400 nm. The corresponding simulated results also demonstrate a vigorous shift in the peak spectral position of transmission as a function of AAAs period [see Fig.\ref{fig:6}(e) and \ref{fig:6}(f)]. Moreover, as the lattice period decreases the transmission peaks broaden and blueshift with maximum efficiency. This shift in the transmission wavelength resulted from the in-phase coupling of SPP-Bloch modes with other surface waves that are excited by the AAAs at Au/quartz interface \cite{Kofke2009}. Similarly, in another study, by varying the size of the lattice constant 'a' of an array of square holes from 250 nm to 650 nm, the peak transmission wavelength redshifts to longer wavelengths and the intensity of the transmission peak decreases due to less interaction between LSP modes of the adjacent holes \cite{Shabani2017}. The periodicity of hole arrays ensures effective coupling of evanescent waves and supports SPP propagation, albeit limited by ultrafast damping. Properly tuned lattice geometry and symmetry are thus critical for achieving maximal transmission with enhanced characteristics features. In the next section, we will explore some more studies to understand the dynamics of EOT in different geometries, instead of hole structure, such as metal gratings.

\subsection{Extraordinary optical transmission from metal gratings}\label{subsec3}
\begin{figure}[h]
\includegraphics[width=0.9\linewidth]{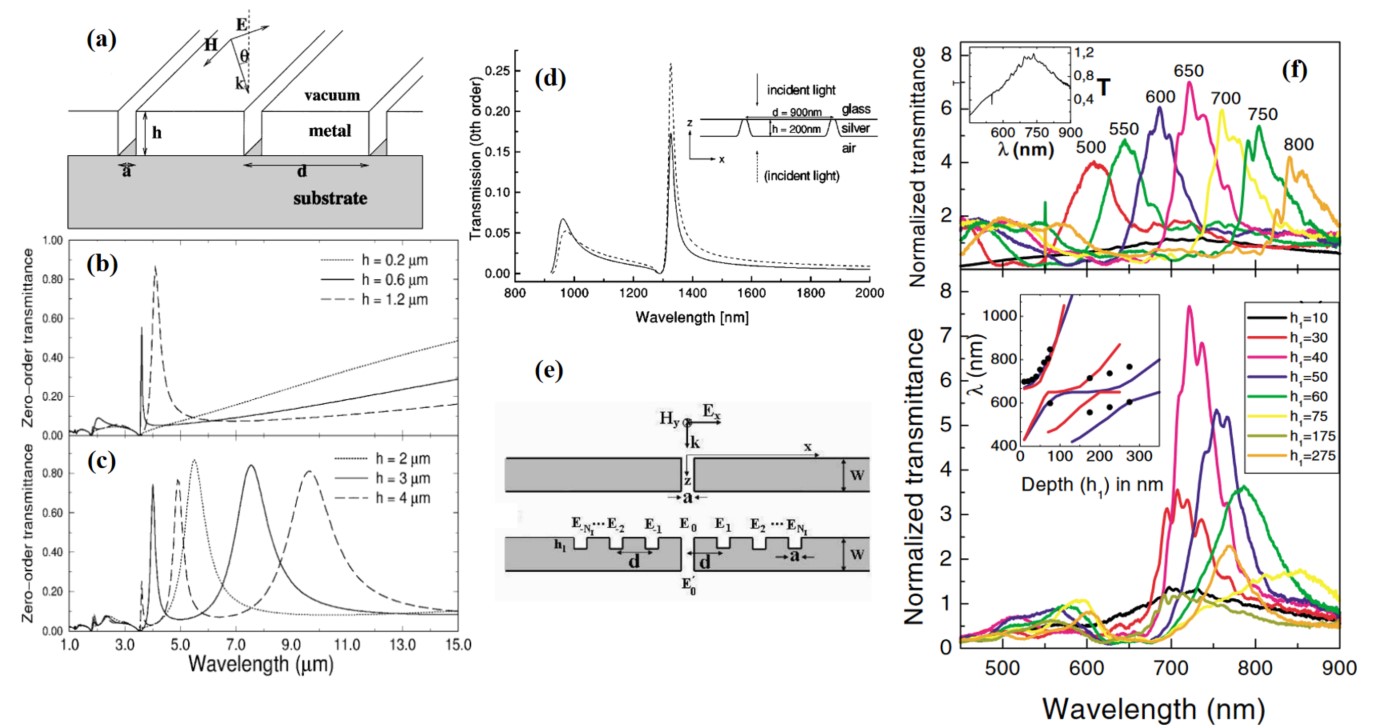} \newline
\centering
\caption{(a) Schematic diagram of lamellar metal gratings with $a = 0.5\mu m$ and $d = 3.5\mu m$. (b,c) Zero-order extraordinary transmission spectra for normal incidence for different values of grating height h. Reprinted (a,b) with permission from \cite{porto1999} Copyright (1999) by the American Physical Society. (d) Zero-order transmission spectra, for normal incidence, from air and glass side, respectively. Reprinted (d) with permission from \cite{Schroter1998} Copyright (1998) by the American Physical Society. (e) Schematic of single slits of width 'a' in a metal film with and without grooves of similar width and period 'd'. (f) Transmission spectra of slit surrounded by $\pm$ 5 grooves (a = 40 nm, W=350 nm, $h_1$= 40 nm) on the input surface, for different array periods d and groove depths $h_1$. Reprinted (e,f) with permission from \cite{Garcia-Vidal2003} Copyright (2003) by the American Physical Society.}
\label{fig:7}
\end{figure}

When compared to hole arrays, metal gratings in the shape of slits are very efficient for transmitting EM radiation under the same conditions. The geometrical parameters such as width, diameter, thickness, and periodicity of grating influence the transmission efficiency in the same way as in the case of holes. In grating-typed structures, upper surface modes couple to lower surface modes as a function of slit height which influences not only the peak intensity but also the spectral position of transmission wavelength. \textit{Porto et al} proposed that transmission resonances are primarily excited by two EM modes in the lamellar metal gratings, as shown in Fig.\ref{fig:7}(a). Through the transfer matrix formalism, it was evaluated that when the peak transmission wavelength is equal to the periodicity of the grating structure, those peaks in the zero-order transmission spectra result from coupled SPP modes [Fig.\ref{fig:7}(b)]. However, when the transmission wavelength is longer than the grating period ($d = 3.5\mu m$), those spectral positions are due to the waveguide resonances, see spectra in Fig.\ref{fig:7}(c). Moreover, the transmittance associated with coupled SPPs modes has a strong dependence on the angle of incidence. In contrast to this, waveguide modes are completely independent of incident angle. 
The waveguide modes are completely different from SPPs modes, for which the slits’ walls play a crucial role in inducing current densities along the walls with opposite signs on both surfaces. These densities follow a parallel path and exit from the other side of the metal without getting affected by the refractive index of the substrate. Since the grating slits permit the coupling of each EM mode. Therefore, modulating the slit’s diameter changes the bandwidth of transmission resonance provided by SPPs. If the diameter of the slit is large (small), the transmitted resonance has a broad (narrow) linewidth in the constituent spectrum. For large slit' widths, the transmission resonances are associated with the waveguide modes, and for this case, the linewidth of the transmission peak gets broader as a function of slit diameter \cite{porto1999}. Another study also indicted the importance of slit width in grating structure for enhanced EOT efficiency. \textit{Iqbal et al} proposed that slit width and its profile play a vital role in grating structure to achieve maximum EOT corresponding to the optimum value of coupling efficiency. Smaller widths of slits offers higher scattering as compared to scattering from the slits with wider widths. Higher scattering yields high efficiency of coupling incident light to SPPs modes due to higher Fourier components. The smaller slits support higher order plasmon modes which start disappearing with the increase in slit width.  Moreover, it was suggested that an optimum transmission from grating device can be achieved for the slit width between one-third to one-half of the 770 nm period of grating \cite{Iqbal2016}. This leads to an optimum slit width range for higher transmission efficiency.\\
The transmission efficiency is also strongly influenced by the choice of the interface above and below the EOT structure. \textit{Schröter et al} reported the difference in transmission intensity due to the excited SPPs modes for different direction of incidence (inset in Fig.\ref{fig:7}(d)). In figure.\ref{fig:7}(d), the transmission spectra show a more intense transmission signal from the glass side in contrast to the air side. This strong transmission results from the mutual contribution of SPP modes, the tunneling photons, and the new avalanche modes that excite at the edges of the slit on the air side and propagate through the cavity towards the glass side \cite{Schroter1998}. 
It should be noted that transmission resonance is very sensitive to the dielectric medium or substrate attached to the part of the metal surface from which light travels to the far field. So if due to substrate the energies of two SPP modes on the upper and lower surface do not coincide, the coupling is not appreciable, and as a result, decoupling is accompanied by a reduction of transmission resonances with two nondegenerate modes appearing in the transmission spectrum with different wavelengths and low efficiencies. This reduction in the transmission intensity strongly depends on the dielectric permittivity of the medium as well as the height of the slits/hole. \cite{Crouse2005}.   
While coherent coupling of SPP modes stimulates the peak efficiency of transmission resonances, their spatial distribution and propagation direction also determine the enhancement in the transmitted signal. \textit{Garcia-Vidal et al} investigated the multiple paths for resonant modes in a groove-structured single-slit metal film [Fig.\ref{fig:7}(e)]. Three main mechanisms have been suggested that contribute to the transmission efficiency and enhanced it upto two-orders of magnitude even from a single slit, (i) the groove cavity modes (GCM), (ii) the in-phase re-emission from each cavity, (iii) the slit’s waveguide modes. GCM can only be generated when the grooves are perforated on the upper surface of the metal facing the illuminating light and facilitates multiple SPP modes contributing to transmittance as compared to the system without grooves. The combination of GCM and in-phase re-emission superimposes and propagates towards the slot where they encounter the waveguide modes inside the long slit. The resonant coupling of GCM and in-phase re-emission from grooves adds an extra peak at 560 nm in the transmission spectrum (not shown here). In Fig.\ref{fig:7}(f), normalize transmittance spectra show that the groove period and its depth strongly modulate the spectral position and peak intensity of transmittance, respectively \cite{Garcia-Vidal2003}. This quantifies that the peak position of the transmission in both hole arrays and grating structures is associated with the periodicity and index of refraction. Whereas the slit’s width and height control the linewidth and intensity of the transmission resonance effectively. In the next section, we will discuss the impact of hole geometry on transmission efficiency. 
\subsection{Effect of hole shape and size on EOT}\label{subsec4}
\begin{figure}[h]
\includegraphics[width=0.9\linewidth]{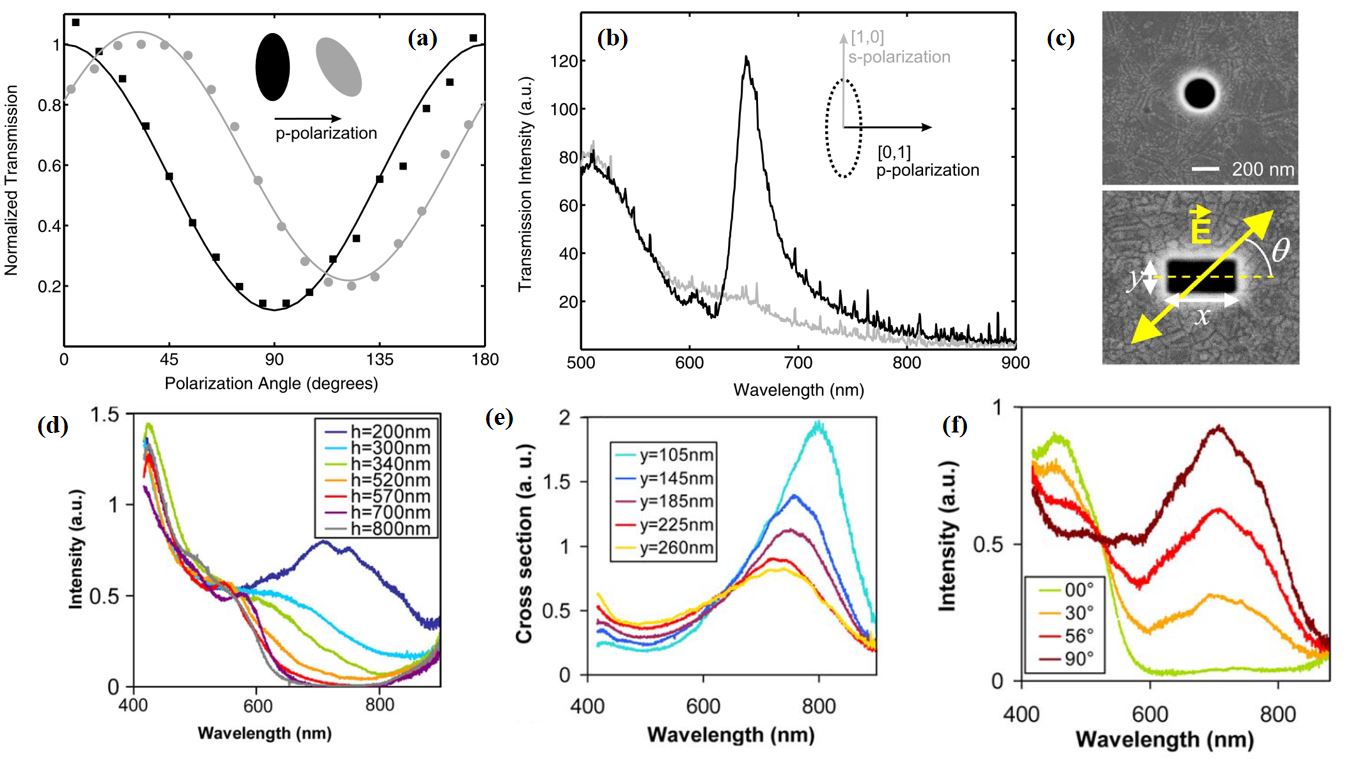} \newline
\centering
\caption{(a) Transmission spectrum through elliptical hole for two orthogonal linear polarization with 0.3 aspect ratio (b) Normalized transmission at (0,1) resonance peak for different orientations of elliptical holes. Reprinted (a,b) with permission from \cite{Gordon2004} Copyright (2004) by the American Physical Society. (d,e) SEM image of the isolated subwavelength hole of diameter d=270 nm, in a silver film, x(y) is the longitudinal (transverse) direction of the rectangle. Transmission spectra for a range of hole depths h. (f-h) SEM image of a rectangular aperture in a suspended Ag film. EOT spectra of various polarization (h=700 nm, x=310 nm, y=210 nm). Transmission spectra for various transverse dimensions Reprinted from \cite{Degiron2004}, Copyright (2004), with permission from Elsevier.}
\label{fig:8}
\end{figure}

The optical characteristics of transmitted light such as polarization, directionality, and nonlinearity greatly influenced by the hole shape and its orientation. \textit{Gordon et al} investigated the transmission properties of elliptical nanohole arrays which provide a mechanism for polarization selectivity in nanophotonic devices. It was shown that the polarization of transmitted light changes drastically for different orientations of elliptical holes and maximum transmission occurs for polarization angle perpendicular to the major axis of the ellipse as shown in Fig.\ref{fig:8}(a). Moreover, the transmission intensity from elliptical holes gets maximum for p-polarization in contrast of s-polarization of incident light [see Fig.\ref{fig:8}(b)].
When the SPP mode propagates parallel to the electric field polarization, the Braggs resonance (0,1) is aligned with the optical polarization which enhances the coupling to the grating parallel to the (0,1) direction. The aspect ratio of elliptical holes also results in the multi-mode transmission due to the coupling of longitudinal and transverse SPP modes. Nevertheless, the increase in the hole diameter decreases the area-to-aperture ratio which limits the propagation path of SPP mode and increases its decay rate, resulting in a blue shift in the transmitted light \cite{Gordon2004}. Because SPPs does not propagate in the cavities, a small spatial region for surface wave propagation incorporates the fast decay of SPPs modes, causing the transmission efficiency to decrease vigorously. On the other hand, the boundary conditions around the nanohole play a crucial role both in the aperture near-field distribution and far-field scattering. When light with appropriate polarization impinges on a single hole, it excites LSP on the ridges of the aperture, as shown in Fig.\ref{fig:8}(c). Because of its dipole nature, LSP causes unidirectional tunneling of SPPs from the hole. The space between the edges plays a significant role in the tunneling and coupling of evanescent waves between both sides. The transmitted light is mainly re-scattered at the ridges through the LSP mode. The increase in the transmission intensity by reducing the hole size is similar to the increase in the scattering cross-section calculated for small metal nanoparticles. The presence of dipole modes on the exit side restricts the spatial dispersion of light and keeps it unidirectional rather diffracting. In addition, the LSP dipole at the ridges causes the re-scattering of the propagating modes and with reduced size, the transmission intensity increases in the same way as the scattering cross-section evaluated for small nanoparticles. In this way, smaller hole diameters confine surface plasmon modes effectively, resulting in sharper transmission peaks.\\ 
\textit{Degiron et al} investigated the role of different shape of holes on the EOT resonance. For a spherical shape hole the transmission spectra demonstrate two peaks for different hole depths 'h' and it was found that for the shallowest holes (h$=200$ nm), the lowest energy peak is supported by LSP mode, as shown in Fig.\ref{fig:8}(d). For a rectangular aperture two exciting modes contribute to transmission and their mutual coupling result in higher and lower energy peaks in the EOT spectrum which correspond to transverse and longitudinal LSP modes, respectively [see Fig.\ref{fig:8}(e)]. As the value of y decreases, longitudinal SPP mode redshifts to longer wavelengths. Nevertheless, the peak transmission normalized to the hole area increases as the hole size is reduced. This is because the open space between the hole edges allows more tunneling and evanescent coupling between the two sides. Moreover, the spatial characteristics of LSP strongly depends on the polarization of incident light and the geometry of the hole. For different polarization angles, the transmission intensity shows a strong angular dependence and get maximum when angle is perpendicular to long axis (x) of the rectangular hole, as shown in Fig.\ref{fig:8}(c).   
Figure.\ref{fig:8}(f) demonstrate the modulation in LSP mode intensity at the ridges by changing the angle between the electric field and longitudinal direction x. When the electric field polarization is perpendicular ($90^\circ$) to the longitudinal axis, propagating mode switches to LSP dipole mode around 700 nm. Only the LSP mode oscillates along the axis normal to the polarization due to its magnetic dipole nature \cite{Degiron2004}. Generally, the number of plasmon modes depends on the spatial orientation and size of each ridge. For a hole with identical edges such as square and cylindrical aperture, the LSP oscillating along each ridge are the same, as a result, transmission is independent of polarization. Whereas, for triangular or rectangular shape holes, the optical properties of EOT light change due to non-centrosymmetric hole structure which causes the excitation of asymmetric SPPs modes on the surface. In this way, polarization sensitivity of non circular holes offers additional control over the transmitted signal. Moreover, the coupling of multi-resonant LSP modes at the ridges with SPPs yields second-and third-harmonic generation at the metal-hole array with high power efficiencies \cite{Kottmann2000, Drobnyh2020}. 
\subsection{Impact of incident light on extraordinary optical transmission}\label{subsec5}
The spatial and spectral characteristics of transmitted light can also be modulated through the exciting pulse characteristics. For instance, the angle of incidence effectively modulates the propagation direction and intensity of the transmitted light. For a minimum deviation in the incident angle, the transmission peak intensity shifts and splits into two new peaks which move in opposite directions and disappear for $\theta = 24^\circ$ \cite{Ebbesen1998, Schroter1998}. Moreover, an oblique incidence results in high transmission efficiency with a much larger cut-off than a normal incidence \cite{Fu2010}. This is due to the fact that SPPs excitation at oblique incidence yields efficient unidirectional propagation and higher relative coupling efficiency in the grating structures \cite{Pisano2018,Yong2006}. In the near-field analysis of surface waves through scanning near-field optical microscope (NSOM), a non-uniform light distribution was observed over the hole array structure. This non-uniformity results from the finite size of the array or due to edge effects. The edge effect breaks the periodicity and SPPs reflected from each edge interferes with other propagating SPPs in between the holes. Another striking fact is that beam polarization causes an asymmetric pattern of near-field along the x and y directions of the periodic structure which results in the asymmetric transmission of light from the hole arrays \cite{Mrejen2007}. The back-scattering of surface waves from the edges of the holes also contributes to the asymmetric patterns \cite{Bravo-Abad2006}.

\subsection{Multimode extraordinary optical transmission}\label{subsec6}
\begin{figure}[h]
\includegraphics[width=0.9\linewidth]{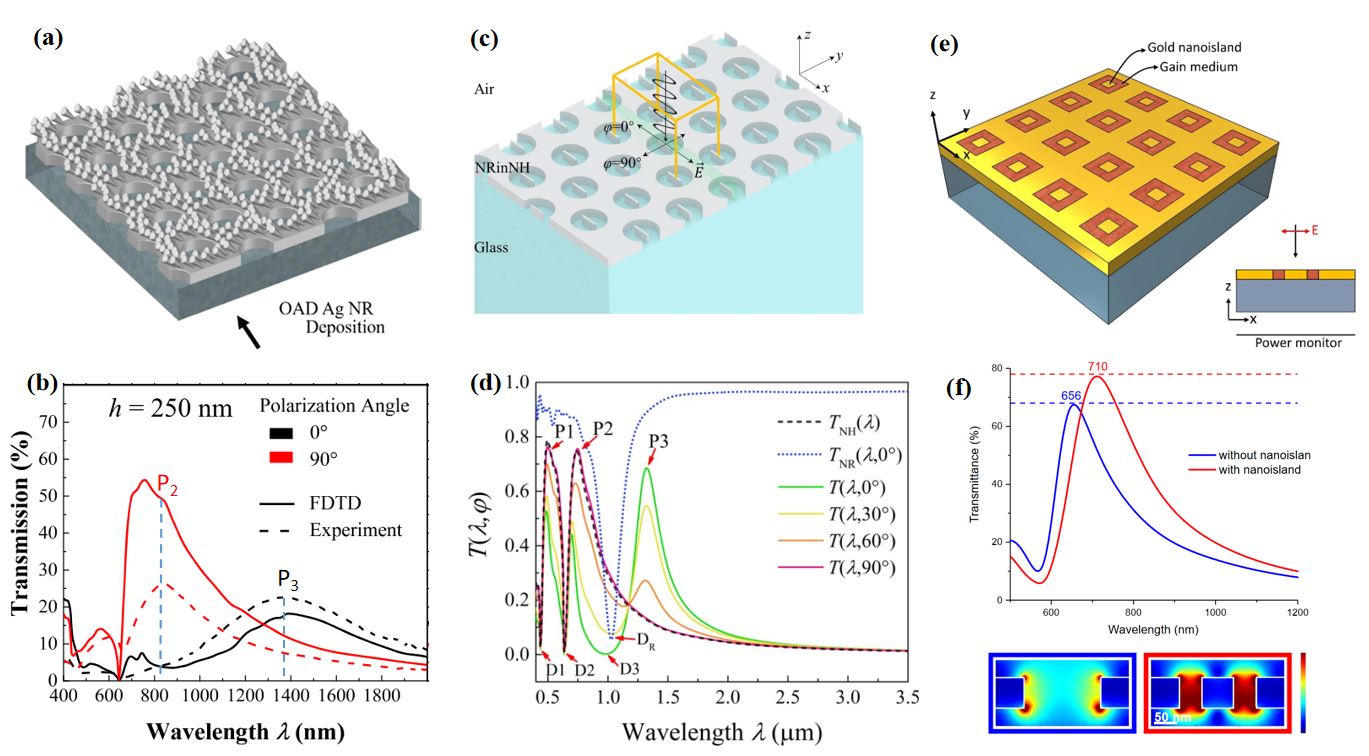} \newline
\centering
\caption{(a) Schematic of Ag nanorods on nanoholes NRonNH in Ag film. (b) Transmission spectra at $\phi = 0$ and $90^o$ for NRonNH with $h = 250$ respectively. Reprinted (a,b) with permission from \cite{Larson2019}. Copyright {2019} American Chemical Society. (c) Schematic diagram of nanorods inside the circular nanoholes array (NRinNH). (d) Transmission T$(\lambda, \phi)$ spectra of the NRinNH (l $= 250$ nm, D $= 340$ nm, L $= 500$ nm, h $= 90$ nm) for different $\phi$. Reprinted (c,d) from \cite{Wang2020} with permission of IOP publishing. (e,f) Schematic of EOT nanostructure of perforated Au film with gain medium inside the holes. Transmission spectra with and without nanoisland. Transmittance spectra for active pump tuning of gain medium. Reprinted (c,e) from \cite{Yildiz2020} with permission of IOP publishing.}
\label{fig:9}
\end{figure}

If the hole cavities are decorated with an auxiliary object supporting localized plasmon modes, then the multi-structured metal nanohole arrays yield multi-mode extraordinary optical transmission. The auxiliary objects in the form of nanorods (NRs) or nanoparticles (NPs) of different shapes situated inside or around the apertures, help to concentrate the energy inside the cavities and enhance the coupling efficiency of resonant modes at both metal surfaces. \textit{Larson et al} demonstrated a new polarization-dependent EOT mode by texturing AgNRs around the spherical nanoholes (NHs), as shown in Fig.\ref{fig:9}(a). When the polarization is normal ($\phi=90^o$) with respect to the long axis of NRs ($h=250$ nm), the spectra show normal EOT mode ($P_2$, red curve) in the visible region. However, when polarization is parallel ($\phi=0^o$) to long-axis of NR, a new mode $P_3$ (black curve) appears which is redshifted from $P_2$ in the near-infrared (NIR) regime [see Figure.\ref{fig:9}(b)] \cite{Larson2019}.
This new EOT mode appears due to strong dipole radiation of LSP modes at the tip of NRs placed around the nanohole. Similarly, \textit{Wang et al} demonstrated that by placing NRs inside the NHs facilitates the coupling of two LSP modes, one on the edges of the NRs and the other at the ridges of the NHs, as shown in Fig.\ref{fig:9}(c). The strong coupling of these two excited LSP modes allows SPPs to tunnel through holes without dispersion and also enables coherent coupling of SPPs, resulting in additional EOT peak. Figure.\ref{fig:9}(d) shows transmission spectra of such a NRinNH structure for different polarization of incident light. By tuning the polarization parallel to long axis of injected NRs, a new EOT mode originated at $P_3$ position along with 68$\%$ enhancement in the transmission efficiency \cite{Wang2020}. \textit{Yildiz et al} also demonstrated the enhancement in transmission efficiency upto 80$\%$ by placing nanoislands inside the holes filled with gain medium, as shown in Fig.\ref{fig:9}(e). Filling nanoislands with gain medium also modifies the transmission resonance with a spectral shift of 46 nm due to mode couplings which changes the optical permittivity of the entire system [see Fig.\ref{fig:9}(f)]. Additionally, pumping gain medium actively enhance the transmission efficiency above 100$\%$ \cite{Yildiz2020}. On the other hand, a combined structure of periodic hole arrays in the form of ring resonators (RR), connected ring resonators (CRR), and split ring resonators (SPR) have shown transmission with high-quality factor (Q) by incorporating the coherent coupling of both local and propagating modes \cite{Ozbay2009}. Moreover, inducing spherical or cylindrical nanoparticles (NPs) inside nanoholes allows coherent mode couplings which results in the enhancement of EOT signal efficiency upto $56\%$ and $48\%$, respectively \cite{Du2019}.\\
These studies have suggested that the addition of metal nanostructures inside the nanohole arrays system not only incorporates the excitation of multiple resonant modes but also enables the coherent coupling between modes which enhances the transmission efficiency with broad spectral bandwidth.
The control of plasmon dynamics through light polarization also modifies the optical characteristics of EOT signal such as wavelength and intensity. Nevertheless, these modifications in the EOT sensor are structure dependent or fixed and once the device is fabricated the operational bandwidth cannot be tuned actively and continuously. This limits the device's capability when it is desired to operate at various frequencies. Consequently, changing the structure and geometry requires time-consuming processes such as design, production, and testing stages. Therefore to achieve transient switching or dynamical control over optical frequency signals, a scalable approach is to modulate surface plasmons in the EOT device through external stimuli. Such type of optimal control not only enables on-demand coherent control of EOT light but also governs programmable photonic systems for PIC applications.

\section{Active tunable EOT for sensor engineering and PIC technology}\label{sec4}
The ability to control the response of photonic devices from an external source give them great versatility to employ them in a variety of photonic applications such as photonic integrated sensing, information processing and single photon sources \cite{Kim2020}. One of the most important functionalities of these photonic systems is operational frequency range. The soul purpose of active tuning of photonic devices is to control the transmission frequency according to the desire applications. For this purpose, many studies have utilized active tunable materials in the form of semiconductors, dielectrics and graphene nanostructures to explore active control of EOT signal in a broad spectral range (UV to IR). Metal films with high plasma frequencies perforated with hole arrays proved to be a good candidate for enhanced EOT in the ultraviolet region \cite{Ekinci2007}. On the other hand, a high index engineered 2D metasurface shows tremendous capacity to manipulate the phase, amplitude and polarization of light with high resolution in visible and IR regions. For instance, \textit{Ye Feng et al} demonstrated close-to-unity resonant transmission in the visible regime from silicon nanodics periodically arranged on silica substrate. The enhanced transmission from these metasurfaces results from the complete suppression of resonant reflection due to destructive interference between electric and magnetic dipole resonances supported by silicon nanodics \cite{YeFen2015}. Moreover, some studies have suggested active control of EOT signal in near-infrared (NIR) and THz regime by employing electrically tunable semiconductor materials layered with metal nanohole structures. 
\begin{figure}[h]
\includegraphics[width=0.95\linewidth]{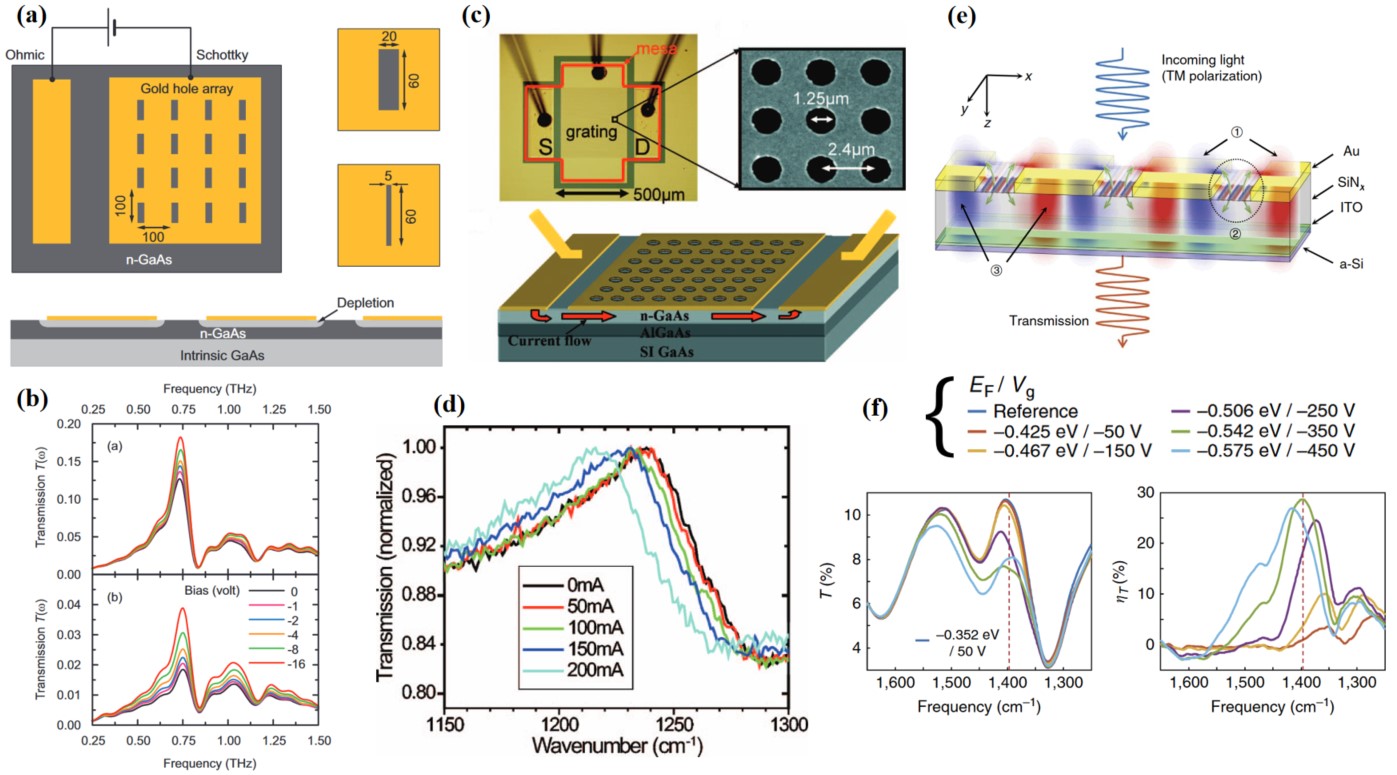} \newline
\centering
\caption{(a) Schematic diagram of metal hole array structure with a cross-sectional view for electrically tunable extraordinary THz transmission. (b) Transmission spectra in THz regime as a function of reverse applied voltage bias from n-GaAs sample structure. Reprinted with permission from \cite{Chen2008} \textcopyright Optica Publishing Group. (c) Schematic of tunable EOT device with n-doped GaAs epilayer and transmission grating with the current path. Optical micrograph of tunable EOT device in mid-infrared regime with source (S) and drain (D) contact. SEM image of transmission grating. (d) Transmission spectra for different current values. Reprinted from \cite{Shaner2007}, with the permission of AIP Publishing. (e) Schematic of subwavelength metallic slit array coupled with graphene ribbons (GPRs). (g) Transmission spectra of GPRs coupled to EOT device for varying graphene Fermi level $E_F$ (or gate voltage $V_g$). (f) Illustration of nanohole structure with graphene beneath the Au nanorod. Transmittance spectra for different values of graphene Fermi Fermi energies $E_F$ \cite{Kim2016} \textcopyright under the terms of Creative Common CC BY license.}
\label{fig:10}
\end{figure}
For instance, \textit{Chen et al} utilized electrically controllable n-doped semiconductor (GaAs) material as a dielectric switch to obtain transmission in the THz regime [see Fig.\ref{fig:10}(a)]. Figure.\ref{fig:10}(b) demonstrate the intensity modulation of THz transmission by applying reverse voltage bias. With the reverse bias the increasing depletion region reduces the losses in dielectric and thereby enhancing the THz transmission resonance. For 5$\mu m$ and 20$\mu m$ wide holes, the intensity modulation depth of the transmitted THz radiation is 52$\%$ and 30$\%$, respectively, which shows that the narrower hole reduce the damping of the resonance yielding a higher modulation depth \cite{Chen2008}. Similarly, \textit{Shaner et al} demonstrated tunable EOT in mid-infrared regime by incorporating electrically tunable n-doped GaAs beneath the metal gratings, as shown in Fig. \ref{fig:10}(c). Figure.\ref{fig:10}(d) shows normalized transmission and shift of 25 cm$^{-1}$ in its spectral position when the current is tuned from 0 to 200 mA. Moreover, transmission strength and linewidth remain stable for high current values which makes the use of such device as tunable mid-IR optical components \cite{Shaner2007}. 
Some other studies have revealed the scope of voltage-tunable graphene nanoribbons (GPRs) for spectral modulation of the transmission signal in the THz regime. \textit{Kim et al} demonstrated efficiency and spectral modulation of transmitted light from EOT structure coupled to graphene plasmonic ribbons, as shown in Fig.\ref{fig:10}(e). By varying the Fermi level E$_F$ of gaphene (or gate Voltage V$_g$) from $-50$ to $-450$ the normalized transmission become stronger and shift to higher energies. Moreover, the modulation efficiency $(\eta_T)$ (normalized transmission spectra with transmission spectrum obtained from bare GPRs device) shows narrow band features that become intense and shift to higher energies [see Fig.\ref{fig:10}(f)] \cite{Kim2016, Gao2021}. Electric-tunable graphene also enables the tuning of weak Raman signals for surface-enhanced Raman spectroscopy (SERS) applications \cite{Gencaslan2022}. Apart from electric control, spectral modulation in the transmission resonances in NIR region has also been demonstrated through an externally applied magnetic field. \textit{Battula et al} investigated the blueshift in transmission with the increase in the magnitude of magnetic field \cite{Battula2007}.
\begin{figure}[h]
\includegraphics[width=0.95\linewidth]{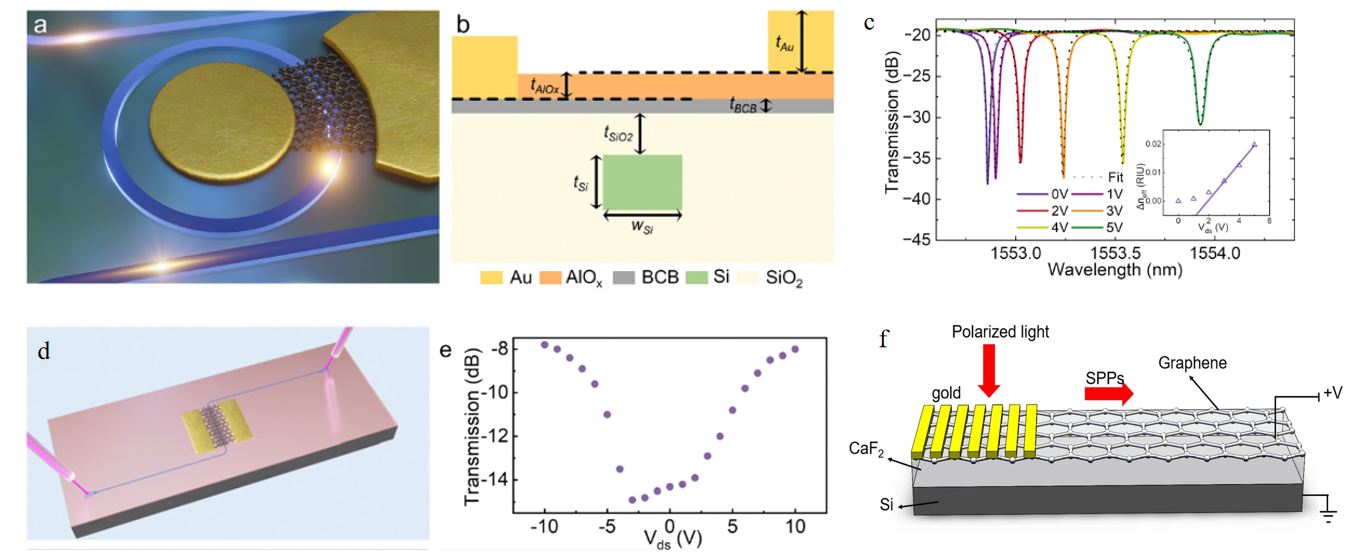} \newline
\centering
\caption{(a) 3D view of hybrid graphene-silicon micro ring thermo-optical (TO) modulator. (b) cross-sectional view. (c) Transmission spectra for various applied voltage bias. Inset shows the change in the refractive index as a function of voltage bias. (d) Graphene electro-absorption (EA) modulator. (e) Transmission spectra at 1550 nm as a function of voltage bias. Used with permission of Royal Society of Chemistry, from \cite{Wu2024}; permission conveyed through Copyright Clearance Center, Inc. (f) Schematic of hybrid silicon-dielectric-graphene-metal grating structure work as a multi-parameter tunable plasmon modulator \cite{Hu2023}. \copyright Creative Commons CC BY license.}
\label{fig:11}
\end{figure}
These studies suggested that electro-optic modulation of plasmon-based photonic devices makes them highly compatible systems to leverage CMOS technology and fabricate hybrid devices for PIC. For instance, the combination of graphene with the silicon on insulators (SOI) platforms offers low cost and large-scale production in contrast to simple silicon-based optical modulators which restrain from low efficiency and speed. One such device is proposed by \textit{Wu et al} as an efficient light modulator for photonic integrated circuits, as shown in Fig.\ref{fig:11}(a). The study demonstrates high-efficiency graphene–silicon hybrid-integrated modulators (cross-sectional view of the device is shown in Fig.\ref{fig:11}(b,c), leveraging gold-assisted transfer to achieve clean interfaces and low contact resistance. The thermo-optic (TO) modulator exhibits a tuning efficiency of $0.037$nm$/$mW, a compact active area of $7.54\mu$m$^2$, and rapid response times [see Fig.\ref{fig:11}(d)]. While the electro-absorption (EA) modulator achieves a bandwidth of $26.8$ GHz, a data rate of 56 Gb$/$s, energy consumption as low as 200 fJ$/$bit and faster switching time (ns) due to high mobility of graphene \cite{Wu2024}. In another study, \textit{Hu et al} proposed a multi-parameter optical modulator to tune optical signal amplitude, wavelength and phase simultaneously in PIC for various applications. Figure.\ref{fig:11}(f) demonstrate the schematic of hybrid silicon-dielectric-graphene-metal grating device. The SPPs modes excited on gratings propagates along the surface of graphene attached to external bias voltage. The change in the Fermi-level of graphene induce a significant increase in mode intensity. Also the wavelength of SPP redshifts when the Fermi-level of graphene changes from $0.3$eV - $0.9$ eV \cite{Hu2023}. Multi-parameter modulation in hybrid silicon-graphene photonic circuits provides on-chip active tuning of several components in PIC and outperform state-of-the-art alternatives in speed, efficiency, and power consumption, making them highly promising for CMOS-compatible, large-scale integrated photonic systems. To get more insight on the role of dielectrics in EOT, we will explicitly discuss some recent studies suggesting broadband EOT resonance and its tunability through dielectric metasurfaces in the following section.

\subsection{Extraordinary optical transmission from dielectric metasurfaces}\label{sec5}
Since its discovery in metal films, extraordinary optical transmission have been a corner stone of nanophotonics and subwavelength optics. While initial studies of EOT focused on metallic structures leveraging surface plasmon polaritons, recent advancements have demonstrated that dielectrics films also support EOT through fundamentally different mechanisms. Unlike metals, dielectric materials such as silicon and silicon dioxide (SiO$_2$) exhibit lower losses, higher stability, and compatibility with existing semiconductor fabrication techniques, making them promising candidates for mid-infrared (MIR) photonic applications.
\begin{figure}[h]
\includegraphics[width=0.9\linewidth]{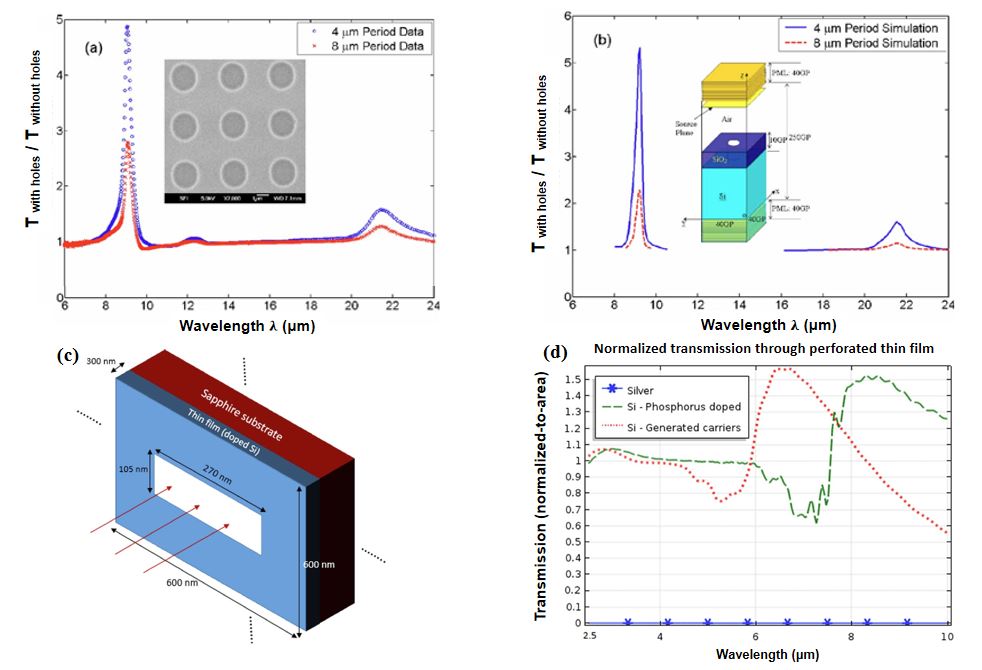} \newline
\centering
\caption{(a) Experimentally measured transmission spectra from silicon dioxide films perforates with 2 $\mu$ m  holes normalized to the transmission from solid film. Inset shows a SEM image of actual sample. (b) FDTD measured transmission spectra for different array periods from the structure sample shown in inset. Reprinted from \cite{Chen2007} with permission from AIP publishing. Transmission spectrum through elliptical hole for two orthogonal linear polarization with $0.3$ aspect ratio (c) Schematic structure of 300 nm thick silicon thin film with a single hole layered on a sapphire substrate. (d) Normalized transmission for different thin films \cite{Mekawey2021}. \copyright Under the terms of Creative Common CC BY license.}
\label{fig:12}
\end{figure}
In contrast to surface plasmon polaritons attributed to enhanced transmission in metal films, EOT from dielectrics is governed by the excitation of surface phonon-polaritons (SPP) which yields transmission resonances in the infrared regime. Recently, Chen et al demonstrated that EOT occurs from thin SiO$_2$ film perforated with subwavelength hole. Figures.\ref{fig:12}(a) and \ref{fig:12}(b) shows experimental and simulated normalized transmission spectra from $1\mu m$ SiO$_2$ film, insets show the sample structures. The SPP resonances in SiO$_2$ only occur over a finite range of frequencies around $9\mu m$ and $21\mu m$ The position of transmission resonance at certain points close to SPP frequencies indicates the strong involvement of SPP in the EOT from SiO$_2$ film \cite{Chen2007}. On the other hand, Hosam et al investigated EOT from silicon thin film perforated with rectangular holes. A schematic of hole structure in 300 nm thick Si film is shown in Fig.\ref{fig:12}(c). 
In this study, the silicon film is either doped or a semiconductor with excess concentration of carriers making it feasible to control Si plasma frequency by tuning the level of excess carriers. The normalized transmission spectra obtained from three different film structure demonstrate an extraordinary transmission at $6.68 \mu m$ for Si-generated carrier film along with peak redshifted to $8.6 \mu m$ for Phosphorous-doped Si film structure. While for silver case, the transmission order is $10^{-4}$ as expected in this range EOT is not possible from silver film because plasmon resonances occur around 900 nm \cite{Mekawey2021}. The controllability mechanism of Si through carrier concentration offers extra degree of freedom to tune device operations and the potential to shift induced plasmon frequencies from visible to the mid-IR spectral range. Moreover, due to low cost fabrication and smoother integration into photonic devices, silicon-based EOT structure can efficiently be replaced with metal-based devices for generating plasmonic platforms in IR spectral region to achieve enhanced vibrational sensing and spectroscopy of molecules.
\begin{figure}[h]
\includegraphics[width=0.8\linewidth]{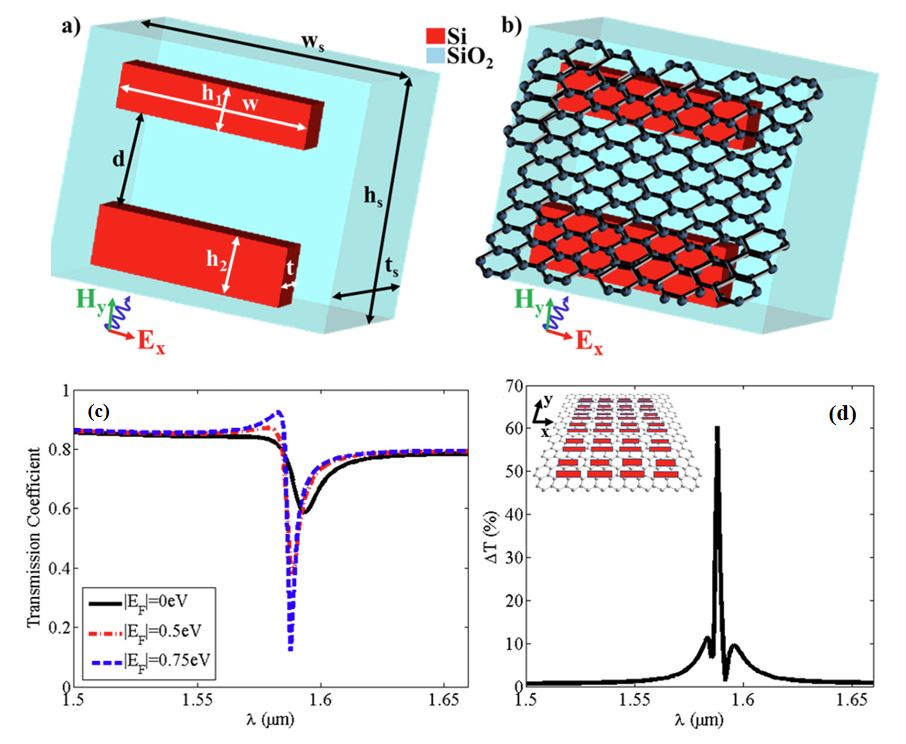} \newline
\centering
\caption{(a) Schematic diagram of periodic arranged Si nanobars and SiO$_2$ metasurface (b) Hybrid structure of all-dielectric metasurface with graphene layer placed over the Si nanobars. (c) Transmission coefficient spectrum as a function of incident wavelength obtained from the hybrid/graphene dielectric metasurface. The transmission amplitude is enhanced for higher doping levels of graphene. (d) Absolute value of transmission coefficient difference for doped and undoped graphene $\Delta T= |T(E_F=0.75$eV$)-T(E_F= 0$ eV$)|$ \cite{Argyropoulos2015}. \copyright Under a Creative Commons license.}
\label{fig:13}
\end{figure}
In another study related to tunable functional response of photonic device, Argyropoulos proposed all-dielectric metasurface with hybrid graphene layer to tune and modulate transmission in near-infrared (NIR) region. A schematic structure of such device is shown in Fig.\ref{fig:12}(e), consisting of a periodic pair of silicon nanobars placed over a silica substrate.
The all-dielectric metasurface is covered with an atomically thick graphene sheet whose properties are adjusted by tuning the Fermi energy levels or chemical potentials through external voltage-bias. The propagating plasmons at the surface of graphene excited in far-and MIR regime can also be tuned through electrical variation in carrier density. The combination of asymmetric nanobars with graphene layer enhances the scattering and transmission responses while overcoming the interband losses of graphene at NIR frequencies. More importantly, in contrast to metallic structures which suffer large Ohmic losses, the hybrid graphene dielectric metasurface offers very low inherent nonradiative Ohmic losses at NIR frequencies which leads to high quality factor Fano and electromagnetic induced transparency making it more compatible for integration in CMOS technology. Figure.\ref{fig:13}(c) show the tunable transmission amplitude for different Fermi energies of graphene layered on all-dielectric metasurface. The system achieves up to 60$\%$ modulation of transmission amplitude  at Fano dip ($\lambda=1.59\mu m $) by tuning graphene's Fermi energy from 0 eV to $0.75$ eV, as shown in Fig.\ref{fig:13}(d) \cite{Argyropoulos2015}. These dielectric platforms offer new opportunities to achieve EOT across a broader spectral range, particularly in the infrared, and exhibit unique advantages such as lower losses, tunability via carrier manipulation, and compatibility with established silicon photonic technologies. This shift to dielectric materials not only broadens the functional range of EOT but also opens new avenues for applications in photonics and sensing.

\section{Ultrashort pulse-driven extraordinary optical transmission}\label{sec6}
From the beginning, many studies has reported EOT phenomenon by illuminating nanohole-metal structure with a continuous plane-wave (CW) source. Nevertheless, some studies have reported the excitation of surface plasmon through an ultrashort light pulse and demonstrated the modulation of spatiotemporal properties of the transmitted signal in real time.  
Ultrashort pulses with broad spectral and temporal functional degrees of freedom offer simultaneous excitation of coherent pathways, and their interference gives rise to transient modulation in plasmonic response \cite{Bahar2022}. In comparison, the possibility of active tuning of ultrafast plasmon dynamics in the steady state through continuous wave (CW) excitation is quite low because of the rapid redistribution of deposited energy through the system in a short timescale \cite{Baumert1997}. Tuning the optical properties of SPP and LSP through ultrashort light pulses brought new possibilities in ultrafast active plasmonics \cite{MacDonald2009}. Moreover, in the evolution of the EOT system, the temporal characteristics (pulse shape, duration, and spectral bandwidth) of the driving pulse are used as the functional degrees of freedom for the time-resolved tuning of SPP and LSP resonances. \\ 
\begin{figure}[h]
\includegraphics[width=0.95\linewidth]{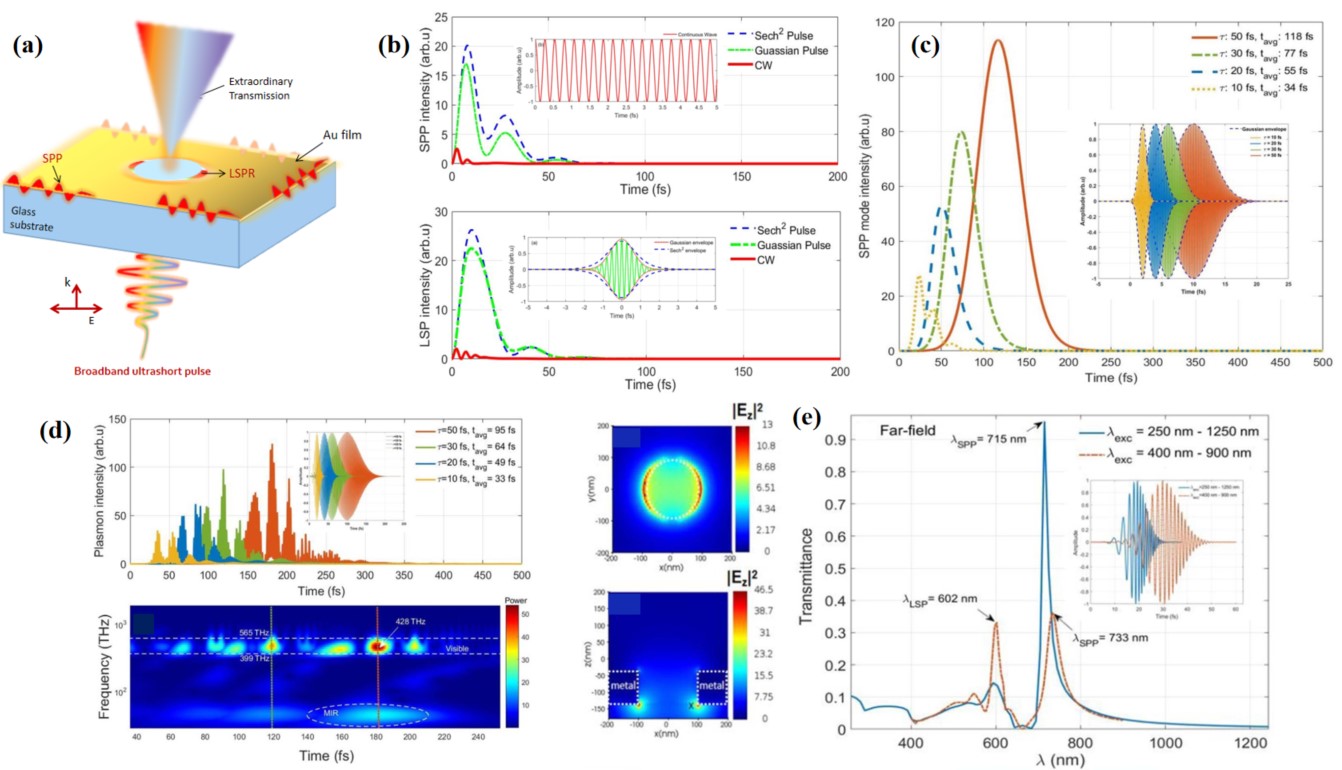} \newline
\centering
\caption{(a) Schematic diagram of EOT device, surface plasmons are excited and probe through an ultrashort light pulse (b) Field intensities of SPP and LSP modes for different pulse shape (inset shows the waveforms of continuous wave (CW) and Gaussian pulse. (c) Plasmon mode intensity as a function of time for various durations of Gaussian pulse, the average lifetime ($t_{avg}$) of SPP increases linearly with pulse width. (d) Power spectra of plasmon field measured through FDTD method for different pulse durations and corresponding CWT spectra. (e) Field distribution at the hole edges and interface of the Au metal film. (f) Transmission spectra for pulse sources with different spectral bandwidths. Used with permission of IOP publishing, from \cite{Asif2023}; permission conveyed through Copyright Clearance Center, Inc.}
\label{fig:14}
\end{figure}
Recently, Asif et al have proposed active control of ultrafast plasmon resonances by exciting the EOT structure through the ultrashort light pulse as shown in Fig.\ref{fig:14}(a). Ultrashort pulse bandwidth modulates the spectral response of surface plasmons and enhances their coherent interferences which yields high transmission efficiency in the visible regime. On the other hand, an ultrashort pulse of sub-100 fs integrates the oscillation energies of plasmon polaritons which results in the extension of the decay time of surface plasmon. The extension in the lifetime of SPP/LSP modes not only enhances EOT signal efficiency but also elongates its coherence time \cite{Asif2023}. Moreover, pulse shape modulates signal intensity and peak linewidth according to the envelope's function. In another study, \textit{Pearce et al}. proposed the modulation of the transmission signal and probing the dynamics of resonant modes by using two identical femtosecond pulses with a time delay of 5 fs. A computational model of perforated Au film of thickness (h) deposited on a $Si_3N_4$ substrate.. The interhole coupling that results from the excitation of LSP is enhanced by the implication of the second pulse. For a time delay of 10 fs between two pulses, the transmission peak corresponding to LSP modes is broadened and intensified significantly. The increase in the intensity and modification in the linewidth results from the additional modes excited at the peak oscillating time of resonating modes. Since the decay time of SPP oscillations is around 20-30 fs, therefore, injecting another pulse after 10 fs modified the bandwidth of oscillating modes and contributed to peak transmission \cite{Pearce2016}. 
\section{Quantum control of extraordinary optical transmission}\label{sec7}
\begin{figure}[h]
\centering
\includegraphics[width=0.9\textwidth]{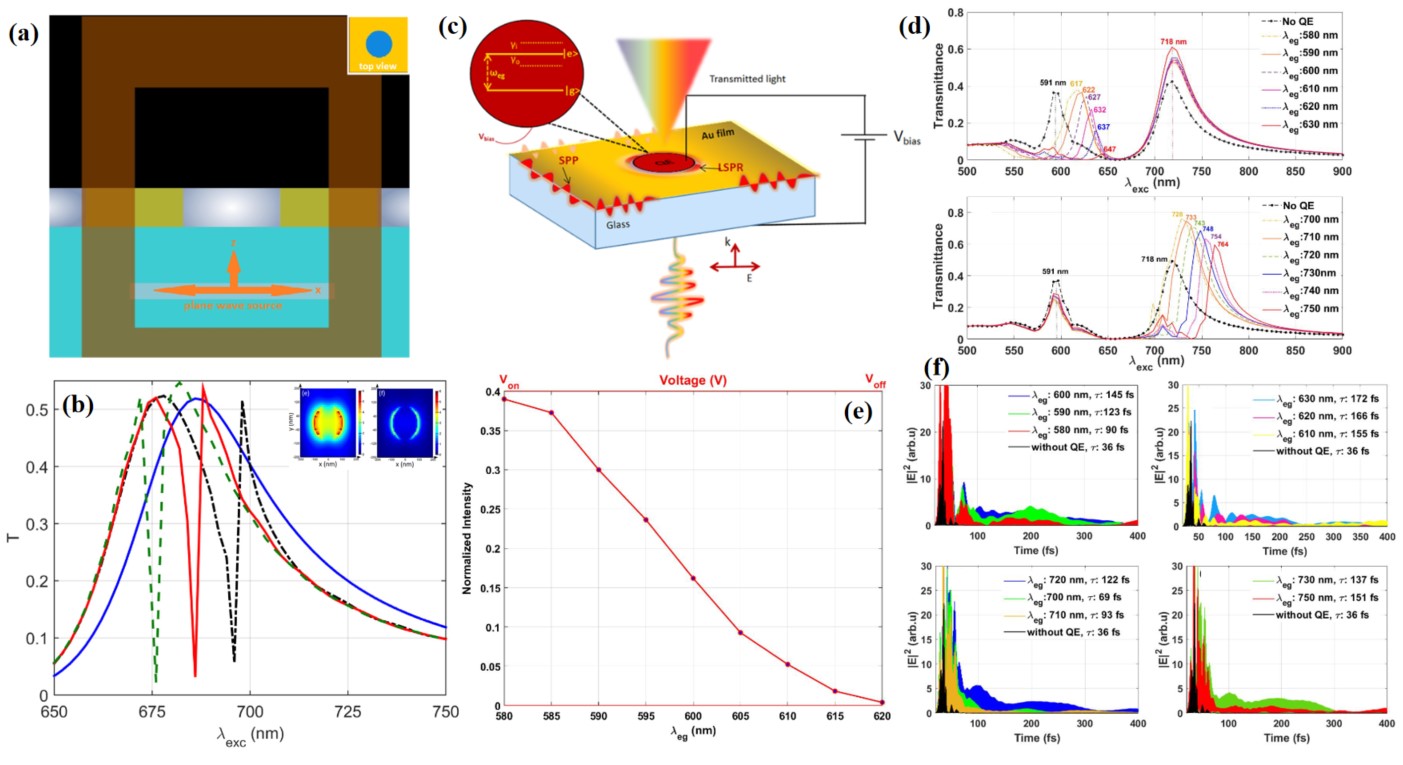}
\caption{(a) Cross-sectional view of a FDTD simulation setup of Au film layered on a glass substrate, excited through a plane wave source. (b) Transmission spectra with and without auxiliary molecule of different level-spacing. inset: Electric field profile in xy plane at the top surface of the Au film for $\lambda=686$ nm incidence. Reprinted from \cite{Sahin2020}, Copyright (2020), with permission from Elsevier. (c) Schematic of the programmable EOT device with a voltage-tunable QE. (d) Spectral shift in the transmission spectra for various level-spacing of QE. (e) Voltage-controlled transmission at a fixed excitation frequency. (f) Power spectra of plasmon modes for different wavelengths of QE and calculated average lifetime. Reprinted (d-f) with permission from \cite{Asif2024} Copyright (2024) by the American Physical Society.}
\label{fig:15}
\end{figure}

The progression in the field of optical transmission through opaque materials has surpassed the limitations of subwavelength optics and nanoscale photonics. Many promising applications have been optimized through the active and passive controlling of EOT sensors, facilitated by hybrid metallic systems designed with subwavelength nanocavity structures. The future challenges concerning the intense and coherent optical transmission from solids are heading toward the quantum control of light through nanocavity-coupled atomic systems.
Controlling EOT signals in real-time and optimizing the EOT structure as a multifunctional in-situ programmable device through active tuning is crucial for the development of quantum technologies such as quantum computing and information processing. For such control, the quantum properties of SPPs and LSP in an EOT device are manipulated by excitons such as quantum emitter (QE) under cavity quantum electrodynamics. The interaction of the quantum objects with resonantly focused light modes constitutes a scalable and on-chip integrated system for probing cavity-enhanced single-photon light sources. As a result, a strong interaction between light and matter can be achieved from such a system governing vacuum Rabi oscillations. This strong light-matter interaction also incorporates optical nonlinearity, which is tunable at a single-photon level with an ultrafast time scale. To manifest a cavity-controlled EOT light, Sahin et al. proposed coherent control of plasmon resonances in the EOT sensor by embedding an auxiliary molecule inside the nanohole structure of a metal film. The coherent control of SPPs/LSP modes is mediated by \textit{path interference effects}. In the FDTD simulation method, an optically thick Au film with subwavelength holes standing on a glass substrate is excited through a plane wave source. A cross-sectional view of a single-hole structure with a molecule inside the cavity is shown in Fig.\ref{fig:15}(a). The excitation of SPP is accompanied by the localized modes around the ridges of spherical holes. The dynamics of excited SPPs at the metal-glass interface and LSP around the edges are evaluated in the steady-state regime through the coupled harmonic oscillator model and the energy exchange between resonant oscillating modes has been scrutinized in the light of Fano resonance \cite{Sahin2020}. In addition, the suppression of the EOT signal is observed at the level-spacing of QE which shows the coherent exchange of quantum energies between two oscillating systems analogous to electromagnetic-induced transparency [see Fig.\ref{fig:15}(b)]. \\  
Placing QE inside the nanohole of the EOT structure not only facilitates coherent control of plasmon modes but also probes high-density field regions through strong coupling which regulates the beam profile without changing the geometrical properties of the system. The strong QE interaction also induces hybrid polaritonic states which incorporate enhanced photon correlation measurements. Many studies have investigated the benefits of QE-plasmon coupling, in the weak and strong regimes \cite{Chang2006, Zhou2019} due to enhanced fluorescence, controlled spontaneous emission, and single photon sources \cite{Chang2006, Zhou2019, Hutchison2011, Akimov2007}. On the other hand, the long lifetime of the quantum oscillator (QE) has shown a remarkable impact on the ultrafast dissipation of the plasmonic field \cite{emre2022}. Some studies have reported the control of dissipation dynamics of LSP modes through QE and probed plasmonic field intensity at a single molecular level \cite{Asif2022, Ovali2021}. In an EOT structure, the hybrid dynamics of a plasmon-emitter system can be modulated by using a spectrally tunable QE system. For spectral tuning in the optical regime, voltage tunable materials such as semiconductors quantum dots can be employed for electric tuning \cite{Flatte2008, Wen1995, Empedocles1997}. Nowadays, a variety of materials such as monolayer WSe2 and TMd with large transition dipole moments are available that offer enhanced electro-optic tunability \cite{Gunay2020, Sie2014, Darlington2023}. Electro-optic tuning of the EOT signal provides a better solution to constitute a scalable and on-chip tunable EOT device and incorporate it into a photonic integrated circuit \cite{Kim2020, Vasa2018}.\\ 
Recently, Asif et al proposed an electrically tunable EOT device operating in the visible regime [see Fig.\ref{fig:15}(c)]. The tunability is governed by a two-level QE whose resonance can be easily modified by applying an external voltage bias \cite{emre2022}. By changing the transition frequency of QE through external voltage, as shown in Fig.\ref{fig:12}(d), the spectral response of hybrid plasmon modes is not only tuned coherently but also the temporal bandwidth is enhanced at a particular spectral shift. The proposed electrically programmable EOT device operates with two main functionalities. (i) It can be operated efficiently at different incident frequencies and the poor EOT signal at a given operational frequency can be enhanced three orders of magnitude by applying a proper bias voltage. (ii) The device can be operated at a fixed frequency $\lambda_{exc}$, but the EOT signal can be electrically tuned continuously [see Fig.\ref{fig:15}(e)]. In addition, the long lifetime of QE helps to enhance the oscillating time of  LSP and SPPs at a particular spectral shift by employing Fano resonance [see Fig.\ref{fig:15}(f)]. Thus, this approach provides an indispensable EOT device that can be operated off-resonantly by coupling plasmon modes with tuned level-spacing of plasmon modes even in a weak coupling regime \cite{Asif2024}. 

\section{Tunable EOT device as a potential component in photonic integrated circuits}\label{sec8}
Sub-wavelength optics and nanophotonics based on plasmonic nanostructures have revolutionized the photonics industry by miniaturizing the bulky optical elements to nanoscale integrated circuits (PIC) \cite{Elshaari2020}. With high optical functionalities, these photonic elements operating on the quantum level hold tremendous potential for sensing, imaging, 5G telecommunication, high-performance computing, information processing, and advanced quantum technologies \cite{Su2023,Badri2021,Gordon2008}. In integrated photonics, the main role of basic photonic components is to generate, split, couple, modulate, polarize, and switch the photonic signal according to the desired application. Mostly, the integrated photonic systems are designed to perform a particular functionality and implemented as application-specific PIC. To achieve optimum performance these chips require several fabrication cycles and any change in the system specifications requires a new design which is costly and time-consuming. Moreover, with an increasing demand for broad frequency bandwidth, high-speed operations, and low power consumption during operations, conventional PIC elements require scalable optical elements that can leverage PIC towards programmable photonic circuits with on-chip tunability, high-performance efficiencies, and broad operational bandwidth across the entire electromagnetic spectrum \cite{Daniel2024}. In this context, plasmon-based EOT devices bring compactness to numerous photonic functions and play a vital role in enhancing light coupling, filtering, beaming, and instantaneous frequency measurements \cite{Yu2013}.
Moreover, due to their enhance light efficiencies they can be used as waveguides, source generator and high speed transmitters \cite{Kim2018}. Also, their active tunable operational bandwidth make them highly compatible as on-chip sensors, filters and ultrafast switches \cite{Kim2020, Rodrigo2016}. In silicon-based photonic circuits, the silicon-dielectrics components suffer with low light confinement. Hybridizing surface plasmons not only enhance the light confinement in conventional dielectric components but also get compensation for higher Ohmic and propagation loses \cite{Sun2021}. Combining both strength and weaknesses of these materials showing promise for scalability and performance challanges for future PIC. 
\begin{figure}[h]
\centering
\includegraphics[width=0.9\textwidth]{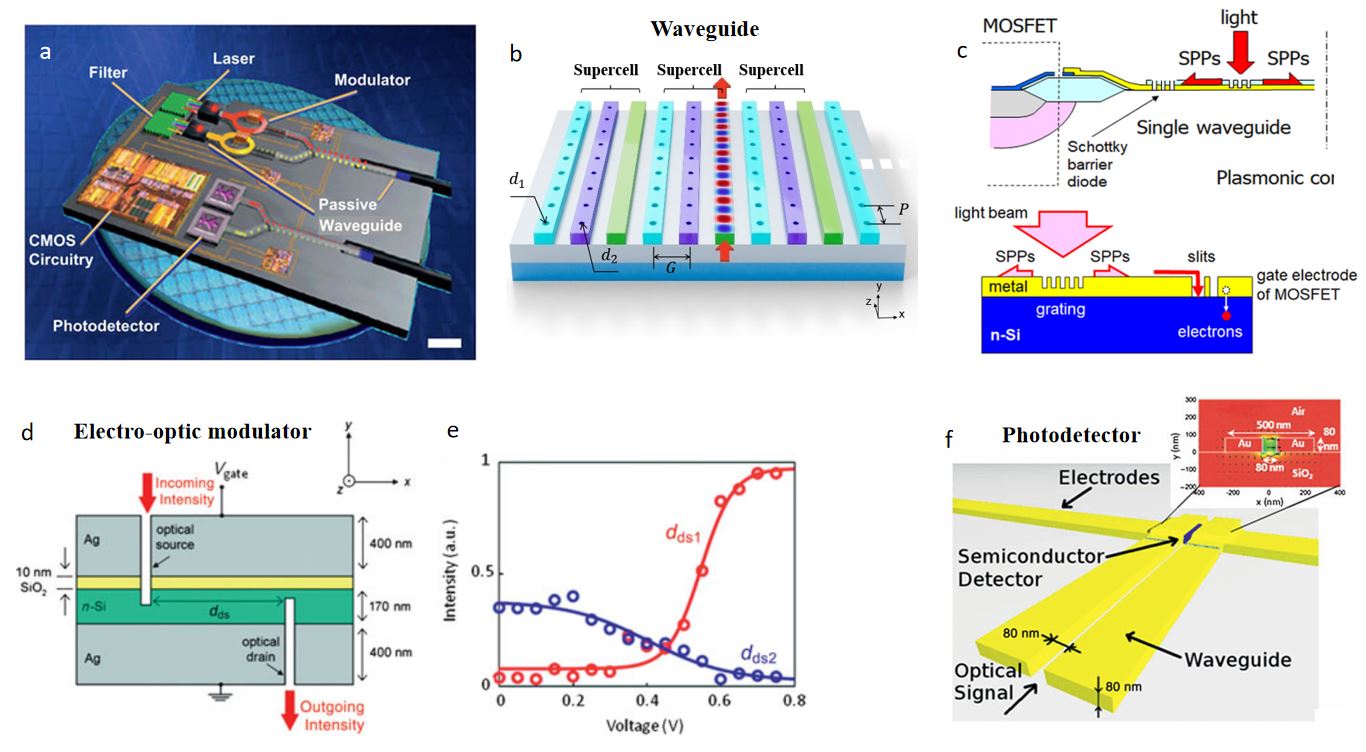}
\caption{(a) Schematic of plasmon-based nanophotonic components on a small wafer footprints for ultracompact and high performance photonic integrated circuit \cite{Sorger2012}. (b) Schematic view of nanohole metamaterial enabled waveguides as a high density waveguide array \cite{Yi2024}. \copyright Under the terms of Creative Common CC BY license.  (c) Plasmonic waveguide for MOSFET chip, a source of light generation and detection for PIC \cite{Fukuda2019}. (d) Plasmonic electro-optic modulator in the form of waveguide integrated silicon nanophotonic device. (e) The modulator yields a 1 dB$\mu$m extinction ratio due to plasmonic MOS mode and the tunable material’s (ITO) ability to modulate its optical loss. (f) Plasmon-based photodetectors for PIC. Reprinted (a, d-f) from \cite{Sorger2012}, Copyright (2012), with permission from Springer Nature.}
\label{fig:16}
\end{figure}
A schematic of photonic integrated circuit chip with high functionality plasmon-based photonic components is shown in Fig.\ref{fig:16}(a) \cite{Sorger2012}. Here, the key features of plasmon-based EOT device that is relevant to PIC technology are source generation, waveguides, electro-optic modulation and photodetection. The novel feature of EOT device is to provide enhance optical light beyond the diffraction limit which gives rise to optical mode for opto-electric devices. For on-chip communication, metamaterial enabled high density waveguides with hole arrays provides extremely strong lateral confinement with less propagation losses [Figures.\ref{fig:16}(b)] \cite{Yi2024}. Metal-insulator-metal (MIM) and Metal-dielectric grating structures also support strong field enhancement with large cut-off and propagation length [Fig.\ref{fig:16}(c)] \cite{Fukuda2019}. For optical modulation, the function of PIC is to switch and route data in the form of electric and optical signal which is called gate transistors. In contrast to all-optical switches or optical transistors which requires higher power laser beams to achieve the necessary gating effect, plasmonic electro-optic modulators (EOM) based on silicon on insulator (SOI) slit waveguide design with active material or active tunable EOT device provide high extinction ratio and indicate a promising path for light-matter enhanced optical switching [Fig.\ref{fig:16}(d) and \ref{fig:16}(e)]. In addition, the electro-optic modulation of such devices gives them extra merit to implement on-chip integration and manipulation for photonic processing units with high-performance output and enhanced optical functionalities \cite{Guo2017}. 
Following the generation, routing and modulating the optical signal, the conversion from optical to electric domain is realized via photodetector [Fig.\ref{fig:16}(f)].\\
To cover all other applications such as filtering, sensing and switching in PIC technologies are beyond the capacity of this review article, since here we concentrated on the fundamental aspects of EOT and the revolutionary progress it has made over the 25 years. In our opinion, the shift from passive to active control of photonic system boosted the current state-of-the-art PIC systems in which a programmable EOT device allows a dynamical control of frequency bandwidth across the optical spectrum and optimizes electro-optic, optoelectronic, and quantum electronic modulation of photonic elements in PIC, photonic communication systems, quantum computing, and information processing technologies \cite{Giordani2023, Ying2016, Genet2003}.
\section{Conclusion}\label{sec9}
Extraordinary optical transmission (EOT) has emerged as a cornerstone phenomenon in subwavelength optics, offering unparalleled control over light transmission through a nanohole array structure and the ability to transmit highly efficient, unidirectional, and coherent optical light that has propelled its use in modern nanophotonics. This review highlights the interplay of geometrical and physical parameters in shaping the EOT phenomenon in perforated metallic and dielectric films. For past two decades extensive research on EOT has established comprehensive insights into the mechanisms and optimization strategies for enhanced transmission. We explored the rich history and advancements in EOT, beginning from its theoretical foundation to its experimental realization, and subsequently, the evolution of passive and active control mechanisms. Passive EOT control, governed by structural parameters such as hole geometry, array periodicity, film thickness and dielectric environment provides a foundational approach to tailoring the optical response of nanohole structures. By systematically varying geometrical and material parameters, it is possible to achieve significant enhancements in transmission efficiency and spectral tunability. While passive designs are fixed post-fabrication, they have demonstrated significant impact in applications requiring high-performance, static optical components, developing static optical filters, wavelength-selective sensors, and subwavelength imaging devices.\\
However, active control has transformed EOT from a static optical phenomenon into a programmable platform for next-generation photonic systems. Active tuning enables real-time tunability of spectral, spatial, and temporal properties of EOT signals. The incorporation of electrically tunable materials such as graphene, doped semiconductors, quantum emitters, and tunable dielectrics into EOT structure, has unlocked the potential for on-demand modulation across a broad spectral range, from visible to terahertz frequencies. These innovations have made EOT devices viable for applications such as photonic switches, sensors, and modulators within photonic integrated circuits (PICs). Similarly, ultrafast optical pulse excitation has extended the temporal coherence and lifetime of plasmonic modes, enabling higher transmission efficiencies and spectral tunability on ultrashort timescales.\\
Nevertheless, active control also presents challenges, including increased complexity in device fabrication and the need for efficient integration with existing platforms. For instance, while graphene-based systems offer unparalleled tunability, their performance can be hindered by interband losses, which require innovative approaches such as hybrid dielectric metasurfaces to overcome. Furthermore, scalability and cost-effectiveness remain key considerations for transitioning active EOT devices from laboratory demonstrations to widespread commercial applications. 
As EOT continues to bridge the gap between nanophotonic and quantum technologies, its role in next-generation optoelectronic and sensing applications is poised to expand. Tunable EOT device with dynamic, high-performance optical functionalities underscores its critical role in advancing PIC technologies, quantum computing, and adaptive sensing solutions. The continued development of active EOT devices promises to bridge the gap between traditional photonic and emerging quantum technologies, paving the way for a new era of reconfigurable and multifunctional photonic devices. The research outlined in this review underscores the need for interdisciplinary collaboration to unlock the full potential of this remarkable phenomenon, ensuring its pivotal role in the advancement of PICs and beyond.

\backmatter

\bmhead{Acknowledgements}

R.S. and H.A. acknowledge support from TUBITAK No. 121F030 and 123F156.

\bibliography{sn-bibliography}

\end{document}